\documentclass[11pt]{article}

\usepackage{graphicx}

\usepackage{amssymb}

\begin{document}
\tolerance=10000

\title{
{\bf Continuous time random walk} \\ 
{\bf and parametric  subordination}\\  
{\bf in fractional diffusion}
}
\author{
Rudolf GORENFLO
\\ Dept. of  Mathematics \& Computer Science, Freie Universit\"at  Berlin,
\\  Arnimallee  3, D-14195 Berlin, Germany
\and
Francesco MAINARDI\thanks{Corresponding author,
 E-mail: {\tt francesco.mainardi@unibo.it}}
 \ and Alessandro VIVOLI
\\ Department of Physics, University of Bologna, and INFN,
\\ Via Irnerio 46, I-40126 Bologna, Italy
}

 \date{May 2007}

\maketitle

\vskip -0.5truecm
\centerline{This paper has been published on} 
\centerline{\bf Chaos, Solitons and Fractals, Vol. 34 (2007), pp. 89-103.}

\begin{abstract}
The well-scaled transition to the diffusion   limit
in the framework    of  the theory of continuous-time
random walk (CTRW)  is presented starting from  its representation
as an infinite series
that points out the subordinated  character of the CTRW itself.
 We treat the CTRW as a combination of a random walk on the axis of
physical time with a random walk in space, both walks happening in
discrete operational time. 
In the continuum limit we obtain 
a  (generally non-Markovian) diffusion
 process governed by a space-time fractional diffusion equation.
The essential assumption is that the probabilities for
waiting times and jump-widths behave asymptotically like powers with
negative exponents related to the orders of the fractional  derivatives.
By what we call {\it parametric subordination}, 
applied to a combination of a Markov process with a positively 
oriented  L\'evy process,
we generate and display sample paths for some special cases.
\end{abstract}

\noindent
{\bf Keywords}:
Parametric subordination, random walks, anomalous diffusion, 
fractional calculus,
renewal theory,
 power laws, L\'evy processes.
\\
{\bf PACS}:  05.40.-a, 89.65.Gh, 02.50.Cw, 05.60.-k, 47.55.Mh
\\
{\bf MSC 2000}:
26A33,  
45K05,  
47G30,  
60G18, 
60G50, 
60G51, 
60J60. 

\newpage
\def\pni{\par \noindent}
\def\vsh{\smallskip}
\def\s{\smallskip}
\def\vs{\medskip}
\def\vvs{\bigskip}
\def\vvvs{\bigskip\medskip} 
\def\vsp{\par}
\def\vsn{\vsh\pni}
\def\cen{\centerline}
\def\ra{\item{a)\ }} \def\rb{\item{b)\ }}   \def\rc{\item{c)\ }}
\def\eg{{\it e.g.}\ } \def\ie{{\it i.e.}\ }
\def\sg{\hbox{sign}\,}
\def\sgn{\hbox{sign}\,}
\def\sign{\hbox{sign}\,}
\def\e{{\rm e}}
\def\exp{{\rm exp}}
\def\ds{\displaystyle}
\def\dis{\displaystyle}
\def\lan{\langle}\def\ran{\rangle}
\def\lt{\left} \def\rt{\right}  
\def\lra{\Longleftrightarrow}
\def\d{\partial}
\def\dr{\partial r}  \def\dt{\partial t}
\def\dx{\partial x}   \def\dy{\partial y}  \def\dz{\partial z}
\def\rec#1{{1\over{#1}}}
\def\zr{z^{-1}}
\def\hatt{\widehat}
\def\epsilons{{\widetilde \epsilon(s)}}
\def\sigmas{{\widetilde \sigma (s)}}
\def\fs{{\widetilde f(s)}}
\def\Js{{\widetilde J(s)}}
\def\Gs{{\widetilde G(s)}}
\def\Fs{{\wiidetilde F(s)}}
 \def\Ls{{\widetilde L(s)}}
\def\L{{\mathcal L}} 
\def\F{{\mathcal F}} 
\def\NN{{\bf N}}
\def\RR{{\bf R}}
\def\CC{{\bf C}}
\def\ZZ{{\bf Z}} 
\def\I{{\cal I}}  
\def\D{{\cal D}}  
\def\erf{\hbox{erf}}  \def\erfc{\hbox{erfc}}
\def\uks{{\widehat{\widetilde {u}}} (\kappa,s)}
\def\psikappa{\psi_\alpha^\theta(\kappa)}
\vskip -0.5truecm
\section{Introduction}  

Surveying the literature of the past 15 years we can
observe an ever increasing interest
in modelling {\it anomalous diffusion}
processes, namely in diffusion processes deviating essentially from
Gaussian behaviour which is characterized by evolution of the second
centered moment like the first power of time.
The reader interested to these processes is referred to
several educational/review papers and books, including
\cite{Balescu 97,GrigoliniRoccoWest 99,
Hilfer 95a,Hilfer BOOK,HilferAnton 95,
Janicki-Weron 94,MannellaGrigoliniWest 94,
Metzler-Klafter PhysRep00,Metzler-Klafter JPhysics04,
Saichev PhysA05,SaichevZaslavsky 97,
ShlesZaslKlafter Nature93,
Sokolov-Klafter-Blumen_PhysicsToday02,Stanislavsky PRE00,
UchaikinZolotarev 99,West BOOK03,
Zaslavsky PhysRep02,Zaslavsky BOOK05}.

In Section 2, we recall the simplest models  for anomalous diffusion
based on fractional calculus.
They are obtained by replacing in the classical diffusion equation the
partial derivatives with respect to space and/or time by
derivatives of non-integer order, 
in such a way that the resulting Green function
can still be interpreted as a probability density evolving in time
differently from the Gaussian type.

A more general approach to anomalous diffusion is
provided by the so-called {\it continuous time random walk} (CTRW)
 introduced in Statistical Mechanics  by Montroll and Weiss
\cite{MontrollWeiss 65}, see also
\cite{MontrollScher 73,MontrollShles 84,MontrollWest 79,Weiss BOOK94},
which differs from the usual models in that
 the steps of the walker occur at random times
generated by a renewal process.
The sojourn probability density of this  process is known
to be governed by an integral equation and  expressed
in  terms of a relevant series expansion, as it will be recalled
in Section 3.
The concept of CTRW,
can be understood
by  considering a random walk subordinated
to a {\it renewal process}, see \eg \cite{Cox RENEWAL67},
 as pointed out  by a number of authors, see \eg 
\cite{BaeumerMeerschaert 01,
GorMai INDIA03,
Kotulski 95a,
M3 PRE02sub,
Sokolov PRE01a,Sokolov PRE01b,Sokolov PRE02}.

It is well known that the {\it space-time fractional diffusion} (STFD)  equation
and its variants,
including the fractional Fokker-Planck equation,
can be derived from the CTRW
integral  equation, see \eg
\cite{Barkai ChemPhys02,Barkai PRE00,
Hilfer PhysA03,
Metzler PRE98,Metzler-Klafter PhysRep00,Metzler-Klafter JPhysics04,
Saichev PhysA05,
Sokolov PRE01b,Sokolov PhysA01,SokolovKlafter EINSTEIN05}, and references therein.
More rigorously the  passage  from  CTRW to STFD
 can be carried out   via a properly scaled transition to the diffusion
limit (under appropriate
assumptions on waiting times and jumps),
as shown in \cite{UchaikinSaenko 03} and
in a number of papers of our research group, see \eg
\cite{GAR Vietnam04,GorMai INDIA03,GorMai CARRY04,
Scalas PRE04}.

In this paper,  we offer another  scheme
of well-scaled transition to the diffusion limit, a scheme
based on a   modified  concept of subordination that we call
{\it parametric  subordination}. 
To this purpose we lay open 
our general view of subordination in stochastic processes in Section 4.
Then in Section 5, starting from the series expansion
of the sojourn probability density in CTRW we arrive in a well-scaled limit process 
at the relevant {\it integral formula for subordination}.

Finally,    we  consider the problem of how to construct
the sample paths for the STFD based on the above  diffusion limit
of the CTRW.
In Section 6, 
we explain what we mean by parametric subordination whereas 
in Section 7, we describe the numerical procedure and provide sample paths 
for four case studies. The main conclusions are drawn in Section 8. 
\section{The space-time fractional diffusion}

 We begin by considering the Cauchy problem for the
(spatially one-dimensional) {\it space-time fractional diffusion equation}
$$  {\, _t}D_{*}^{\, \beta }\, u(x,t)
 \, = \,
 {\, _x}D_{ \theta}^{\,\alpha} \,u(x,t)\,,
\quad  u(x,0) = \delta (x)\,, \quad x \in \RR,\quad t \ge 0\,, \eqno(2.1) $$
where 
   \{$\alpha \,,\,\theta\,,\, \beta $\} are real parameters
 restricted to the ranges
$$ 0<\alpha\le 2\,,\quad  |\theta| \le \min \{\alpha, 2-\alpha\}\,,
  \quad 0<\beta\leq 1\,.\eqno(2.2) $$
Here
${ \,_t}D_*^{\,\beta}  $
denotes  the
{\it  Caputo fractional derivative}
of order $\beta $, acting on the time variable $t$,
and  $ {\,_x}D_{\,\theta}^{\,\alpha}$
denotes
the {\it  Riesz-Feller fractional derivative}
of order $\alpha $ and
skewness $\theta$,   acting on the space variable $x$.
Let us note that the solution $u(x,t)$  of the
 Cauchy problem (2.1), known as the {\it Green function} or fundamental solution
of the space-time fractional diffusion equation,
is a
probability density in the spatial variable $x$, evolving in time $t$.
In the case $\alpha =2$ and $\beta =1$  we recover
the standard diffusion equation for which the fundamental solution
is the Gaussian density with variance $\sigma ^2 =2t$.

Writing, with $\hbox{Re} [s] > \sigma _0$, $\kappa \in \RR$, the transforms of Laplace and Fourier as
$$     {\L} \lt\{ f(t);s\rt\}=  \widetilde f(s)
 := \int_0^{\infty} \e^{\ds \, -st}\, f(t)\, dt\,,$$
 $${\F} \lt\{g(x);\kappa\rt\}=  \widehat g(\kappa)
  := \int_{-\infty}^{+\infty} \!\! \e^{\,\ds i\kappa x}\,g(x)\, dx\,,
$$
we have the corresponding transforms
of ${ \,_t}D_*^{\,\beta} f(t)  $
and       $ {\,_x}D_{\,\theta}^{\,\alpha} g(x)$ as
$$   {\L} \lt\{  {\,_t}D_*^{\,\beta}\, f(t)\rt\} =
    s^{\, \ds \beta}\,  \widetilde{f} (s)- s^{\, \ds \beta -1}\, f(0)\,,
\eqno(2.3)$$
$$   {\F} \lt\{ {\,_x}D_{\,\theta}^{\,\alpha}\, g(x )\rt\} =
  - |\kappa |^{\,\ds \alpha}  \, i ^{\,\ds \theta \,\sgn \kappa} \,
   \widehat {g}(\kappa )\,. \eqno(2.4)$$
Notice that
$i ^{\,\ds \theta \,\sgn \kappa}= \exp [ i \,(\sgn \kappa)\,\theta\,\pi/2 ]$.
For  the mathematical details  the interested
reader is referred to
\cite{GorMai CISM97,Kilbas-et-al BOOK06,Podlubny 99} on the Caputo derivative,
and   to \cite{SKM 93} on the Feller potentials.
For the general theory of pseudo-differential operators
and related Markov processes the interested reader is referred
to the excellent volumes by Jacob \cite{Jacob BOOKS}.

For our purposes let us here confine ourselves to recall the
representation in the Laplace-Fourier domain of the (fundamental) solution
of (2.1) as it results from the application of the transforms
of Laplace and Fourier. Using $\widehat \delta (\kappa ) \equiv 1$
we have from (2.1)
 $$  s^{\, \ds \beta}\,\widehat{\widetilde{u}}(\kappa ,s) - s^{\, \ds \beta -1}
    = -|\kappa|^{\, \ds \alpha} \,  i ^{\,\ds \theta \,\sgn \kappa}
   \, \widehat{\widetilde{u}}(\kappa ,s) \,,
$$
hence
$$  \widehat{\widetilde{u}}(\kappa ,s)
    =  \frac{ s^{\, \ds \beta -1}}
{s^{\,\ds \beta} + |\kappa |^{\, \ds \alpha} \,
i^{\,\ds \theta \,\sgn \kappa} }\,.
   \eqno(2.5)$$
For explicit expressions and plots of  the fundamental solution of (2.1)
in the space-time domain
we refer the reader to \cite{Mainardi FCAA01}.
There, starting from the fact  that the Fourier transform
$\widehat{u}(\kappa ,t)$ can be written as a Mittag-Leffler function
with complex argument, the authors
have  derived a Mellin-Barnes integral representation
of $u(x,t)$  with which they have proved the non-negativity
of the solution for values of the parameters
$\{\alpha,\, \theta, \,  \beta \}$ in the range (2.2)
and analyzed the evolution in time of its  moments.
In particular for $\{0<\alpha <2, \, \beta=1\}$ we obtain
the stable densities of order $\alpha$ and skewness $\theta$.
The representation of $u(x,t)$ in terms of Fox $H$-functions
can be found  in   \cite{Mainardi JCAM05}.
We note, however,  that the solution of the STFD Equation (2.1) and
its variants has been investigated by several authors
as pointed out in the bibliography in \cite{Mainardi FCAA01}:
here we refer to some of them,
 \cite{BaeumerMeerschaert 01,Barkai PRE01,M3 PRE02sol,
Metzler-Klafter PhysRep00}, where the connection with the CTRW
was also pointed out.
\section{The continuous-time  random walk}

The name    {\it continuous time random walk} (CTRW)
 became
popular in physics after
 Montroll, Weiss and Scher (just to cite  the pioneers)
in the 1960s and 1970s published a celebrated series
of papers on random walks for modelling
diffusion processes on lattices, see \eg
\cite{MontrollScher 73,MontrollWeiss 65}, and
the book by Weiss \cite{Weiss BOOK94} with  references therein.
CTRWs are rather good and general phenomenological models for diffusion,
including processes of anomalous transport,
that can be understood  in the framework of
the classical renewal theory, as stated
\eg in  the booklet by Cox  \cite{Cox RENEWAL67}.
In fact a CTRW can be considered
as a compound renewal process (a simple renewal process with reward) or
 a  random walk {\it subordinated}
to a simple  renewal process.

Basic notions of the CTRW theory,  that hereafter we briefly recall
for the readers' convenience,
are the master  equation (in integral form)
for the sojourn probability density,
 its Fourier-Laplace
representation (known as the Montroll-Weiss formula)
and its series representation.  

A CTRW  
is generated by a sequence
of  independent identically  distributed ($iid$)
 positive  random  waiting times $T_1, T_2, T_3, \dots ,$
each having the same probability density function
 $\phi(t)\,,$   $\, t>0\,, $ and
a sequence of $iid$ random jumps $X_1, X_2, X_3, \dots, $
in $\RR\,,$ each having the same probability density
$w(x)\,,$ $\, x\in \RR\,.$

Let us remark that, for ease of language, we use
the word density also for generalized functions
in the sense of Gel'fand and Shilov \cite{GelfandShilov 64},
that can be interpreted as probability measures.
Usually the {\it probability density functions} are abbreviated
by  $pdf$.
We recall that $\phi (t) \ge 0$ with $\int_0^\infty \phi (t)\, dt =1$
and  $w(x)  \ge 0$ with $\int_{-\infty}^{+\infty} w (x)\, dx =1$.

Setting
$t_0=0\,,$ $\, t_n = T_1+T_2 + \dots T_n$ for $n \in \NN\,,$
the wandering particle 
makes a jump
of length $X_n$ in instant $t_n$,
so that its position is $x_0=0$ for
$0\le t <T_1= t_1\,,$       and
$x_n =    X_1 + X_2 + \dots X_n\,,$
for $ t_n \le t < t_{n+1}\,. $
We require the distribution of
the waiting times
and  that of the jumps to be independent
of each other.
So, we have a compound renewal process (a renewal process with
       reward), compare  \cite{Cox RENEWAL67}.

By natural probabilistic arguments we arrive at the
{\it integral  equation} for the probability density   $p(x,t)$
(a density with respect to the variable $x$)
of the particle being in point $x$ at instant $t\,, $
see  \eg
\cite{GorMai INDIA03,Gorenflo KONSTANZ01,Mainardi Bonn00,Scalas PhysA05,SGM 00,Scalas PRE04},
$$
   p(x,t) =  \delta (x)\, \Psi(t)\, +
  \int_0^t  \!\!  \phi(t-t') \, \lt[
 \int_{-\infty}^{+\infty}\!\!  w(x-x')\, p(x',t')\, dx'\rt]\,dt'\,,
\eqno(3.1)  $$
 in which  the {\it survival function}
 $$\Psi(t) = \int_t^\infty \phi(t') \, dt' \eqno(3.2)$$
denotes the probability that at instant $t$ the particle
 is still sitting in its starting position
$x=0\,. $
Clearly, (3.1) satisfies the initial condition
$p(x,0) = \delta (x)$.
In the Laplace-Fourier domain
Eq. (3.1)  reads as
  $$ \widehat{\widetilde p}(\kappa ,s)
 =   \widetilde {\Psi}(s) +  \widehat w(\kappa )\, \widetilde \phi(s)
   \widehat{\widetilde p}(\kappa ,s)\,,   $$
and using $  \widetilde {\Psi}(s)  = {(1-\widetilde\phi(s)) /s}\,,$
explicitly
 $$
   \widehat{\widetilde p}(\kappa ,s)   =  {1-\widetilde\phi(s)  \over s}
 {1 \over 1- {\widehat w}(\kappa )\,{\widetilde \phi}(s)}\,.
 \eqno(3.3) $$
This Laplace-Fourier representation   is known in physics as the
the {\it Montroll-Weiss equation},
so named after the authors, see \cite{MontrollWeiss 65},
who derive it in 1965 as the basic equation for the CTRW.
By inverting the transforms one can
    find the evolution $p(x,t)$  of the sojourn density for
    time $t$    running from zero to infinity.
In fact, recalling that $|\widehat w(\kappa)| < 1$ and
$|\widetilde\phi (s)| < 1$,
if $\kappa \not= 0$ and
$s \not= 0$, Eq. (3.3) becomes
$$
\widetilde{\widehat p}(\kappa, s) =
\widetilde \Psi(s)\, \sum_{n=0}^{\infty}
[\widetilde \phi (s) \, \widehat w(\kappa)]^n =
 \sum_{n=0}^{\infty} \widetilde v_n(s)\, \widehat w_n (\kappa)
 \,,
\eqno(3.4) $$
and 
we promptly obtain the
{\it series representation of the continuous time random walk},
see \eg
 \cite{Cox RENEWAL67}  (Ch. 8, Eq. (4))
or \cite{Weiss BOOK94}  (Eq.(2.101)),
$$
p(x,t) = \sum_{n=0}^{\infty} v_n(t)\, w_n (x)
=    \Psi(t)  \, \delta (x) + \sum_{n=1}^{\infty} v_n(t)\, w_n (x)\,,
\eqno(3.5)$$
where the functions  $v_n(t)$
and $w_n(x)$ are obtained by repeated convolutions
in time  and in space,
$ v_n(t) = (\Psi * \phi^{*n})(t)$,
and $\, w_n(x) = (w ^{*n})(x)$, respectively.
In particular,
$ v_0(t) = (\Psi*\delta) (t)= \Psi(t)$, $\, v_1(t) = (\Psi * \phi)(t)$,
$\, w_0(x) = \delta(x), \; w_1(x) = w(x).$
In the R.H.S  of Eq (3.5) we  have isolated
the first singular term  related to the initial condition
$p(x,0) =  \Psi(0) \, \delta (x) =\delta (x)$.
The representation (3.5)
can be found without detour over (3.1) by direct probabilistic reasoning
and transparently exhibits the CTRW  as   a  subordination of a random
    walk to a renewal process:
it can be used as starting point to derive the
Montroll-Weiss equation, as it was originally recognized by Montroll and
Weiss \cite{MontrollWeiss 65}.
Though (3.5), while being an
attractive general formula, is unlikely
to lead to explicit answers to rather simple problems,
we consider it as a basic and useful formula for our analysis,
as it will be shown later on.

A special case of the integral  equation (3.1) is obtained for
the {\it compound Poisson process} where $\phi (t) = m\e^ {-mt}$
(with some positive constant $m$). 
Then, the  corresponding master equation reduces after some
manipulations, that best are carried out in the Laplace-Fourier domain,
  to the {\it Kolmogorov-Feller equation}:
$$
  \frac{\d }{\d  t}\,p(x,t)= -m\, p(x,t)+ m\,\int_{-\infty}^{+\infty}
w(x-x')\, p(x',t)\, dx' \, .
\eqno(3.6)
$$
Then, the solution obtained via the series representation reads
$$p(x,t) = \sum_{k=0}^{\infty} \frac{ (mt)^k}{k!} \,
{\e}^{- mt} \,w_k (x)\,.
\eqno(3.7)
$$
Note that only in this case the corresponding stochastic process
is {\it Markovian}.

\section{Subordination in stochastic processes }

In recent years a number of  papers have appeared
where explicitly or implicitly
subordinated stochastic processes
have been treated in view of their relevance
in physical and financial applications, see e.g.
\cite{BaeumerMeerschaert 01,Barkai ChemPhys02,OEBN EDITOR01,M3 PRE02sub,
Metzler-Klafter PhysRep00,Scalas PRE04,
Sokolov PRE01a,Sokolov PRE01b,Sokolov PRE02,
Stanislavsky PHYSA03,
UchaikinZolotarev 99,Wyss-Wyss 01}
and references therein.
Historically, the notion of subordination was originated by Bochner,
see \cite{Bochner 55,Bochner 62}.

We obtain the process  $X(t)$  of our proper interest in the form
$X(t) = Y(T_*(t))$ by randomizing the time clock of a stochastic process $Y(t_*)$ 
using a new clock  $t = T(t_*)$, the non-decreasing
right-continuous random functions   $t = T(t_*)$  and  $t_* = T_*(t)$
being inverse (in the appropriate sense) to each other.
The resulting process $x= X(t)$ is said
to be subordinated to the so-called  {\it parent process} $Y(t_*)$,
and $t_*$ is commonly referred to as the  {\it operational time}.

Our essential  process  for randomizing time is the process $t= T(t_*)$, 
called by us the {\it leading process}.  
Our view is in contrast to that of Bochner's subordination
adopted by  Feller \cite{Feller 71} and others,
see \eg \cite{M3 PRE02sub,Stanislavsky PHYSA03}, 
who put into  the foreground the inverse process
$t_* =T_*(t)$, which actually is a hitting time or
first passage process,
and after Feller 
often  called the {\it directing process}.

In particular, assuming  $Y(t_*)$ to be a Markov process
with a spatial probability density function ($pdf$)
of $x$, evolving in operation time $t_*$,
$q_{t_*} (x) \equiv q(x, t_*)$, and  $T_*(t)$ to be 
{\it a  process with non-negative,  not necessarily independent,
increments} with $pdf$ of $t_*$ depending on a parameter $t$,
$r_t(t_*) \equiv r(t_*, t)$, then the subordinated
process $X(t) = Y(T_*(t))$ is governed
by the spatial $pdf$ of $x$
evolving with $t$,
$p_t (x) \equiv p(x,t),$  given by the
{\it integral formula of subordination}
(compare with Eq (7.1), Ch. X in \cite{Feller 71}
and with Eq. (3.1) in \cite{Mainardi FCAA03})
$$
p_t(x)=\int_0^{\infty} q_{t_*}(x)\, r_t(t_*)\, {dt_*} \,.
\eqno(4.1)$$
If the parent process $Y(t_*)$ is  {\it self-similar}
of the kind that
 its $pdf$ $q_{t_*}(x)$ is such that,
with a probability density $q(x)$ and a positive number  $\gamma$,
$$
q_{t_*}(x)\equiv q(x,t_*) =
t_*^{-\gamma } \, q \lt(\frac{x}{t_*^{\gamma }}\rt)\,,
\eqno(4.2) $$
then Eq. (4.1) reads
$$
p_t(x)=\int_0^{\infty} q\lt(\frac{x}{t_*^{\gamma }}\rt)
r_t(t_*) \frac{dt_*}{t_*^{\gamma }} \,.
\eqno(4.3)$$
\underbar{Remark}:
Feller \cite{Feller 71} and several other mathematicians,
\eg \cite{Jacob BOOKS,Sato 99}, 
in their  treatment of subordination, 
are  mainly interested
in Markov processes. After explicitly saying (in his Section X.7) that
{\it the subordinated process may happen to be non-Markovian}, 
Feller  immediately turns his attention to the search for conditions 
to be imposed on the directing  process that ensure
the subordinated process to be Markovian like the parent process.
For our processes (see next Section) these conditions are in general
not fulfilled.

\section{Subordination in continuous time random walk}

In fractional diffusion an intuitive understanding can be
gained by formalizing the transition from the series representations
(3.4) and (3.5) of a general continuous time random walk (CTRW),
known in the mathematical literature as a renewal process with reward.
We cannot survey the rich literature on the subject, but let us call
here the reader's special attention to the most recent papers by
Piryatinska, Saichev and Woyczynski \cite{Saichev PhysA05}
 and Sokolov and Klafter \cite{SokolovKlafter EINSTEIN05}.
These authors show in differing ways how fractional diffusion
can be obtained from continuous time random walk. In contrast to these
authors we lay out in all details our method of {\it well-scaled
transition to the diffusion limit}, making explicit the meaning of
{\it long-time wide-space} behaviour.
For the general principle of well-scaledness we
 refer to
\cite{GAR Vietnam04,GorMai INDIA03,GorMai CARRY04,Scalas PRE04}.

In the series representation (3.5) for the CTRW
 the running index $n$ corresponds to the so-called "operational time"
$t_*$ in the subordination formula for a continuous (stable)
process. We will pass in (3.5) to the diffusion limit under
the "power law" assumptions (in the Laplace-Fourier domain)
$$ 1 - \widetilde {\phi} (s) \sim \lambda s^\beta ,\quad
  \lambda >0,  \quad  s \to 0^+\,, \eqno(5.1)$$
$$ 1-\widehat{w}(\kappa ) \sim \mu
|\kappa| ^\alpha \, i^{\ds \, \theta \sgn \kappa }\,,
 \quad  \mu >0\,, \quad  \kappa \to 0\,,
 \eqno(5.2) $$
where $\beta$, $\alpha$ and $\theta$ are restricted as in (2.2).
If $0<\beta<1$ and $0<\alpha<2$, 
 Eqs. (5.1) and (5.2) imply fat (power-law) tails
for the densities  $\phi(t)$ and $w(x)$; otherwise,
for $\beta=1$, Eq. (5.1) implies that $\phi(t)$ has a finite first moment
(e.g. the exponential $pdf$),   and,  for $\alpha=2$, Eq. (5.2) implies
that $w(x)$ has a finite second moment (e.g. the Gaussian $pdf$).
For details we refer  e.g. to \cite{GorMai CARRY04}.

The idea is to treat the series expansion (starting from $n=0$)
in (3.5) as an approximation
to an improper Riemann integral.
Being interested on behaviour in {\it large time}
and {\it wide space} we change the units of measurement
in order to make large time intervals and space distances
appear numerically of moderate size, moderate time intervals and
space distances of small size.
To this aim we replace waiting times $T$ by $\tau T$, jumps $X$ by $hX$,
and then send the positive {\it scaling factors} $\tau $ and $h$ to zero,
observing a {\it scaling relation}
that will become mandatory in our calculations.
For conciseness of our presentation we skip the analytical subtleties of
interchanges of summations and integrations. For a strictly analytical
derivation of our final integral equation of subordination we
recommend \cite{Mainardi FCAA03}.

For the CTRW this means replacing
$\,\phi (t)$ by   $\phi _\tau (t)= \phi (t/\tau )/\tau$,
$\,w (x)$ by  $w _h (x)= w (x/h )/h$,
correspondingly
$\,\widetilde{\phi} (s)$ by  $\widetilde{\phi} _\tau (s)=
 \widetilde\phi (\tau s)$,
$\,\widehat {w}(\kappa)$ by   $\widehat {w}_h(\kappa)=
 \widehat {w(}h\kappa)$.
Decorating (3.5) by indices $h$ and $\tau $ gives
$$ p_{h,\tau }(x,t) = \sum_{n=0}^{\infty} v_{\tau,n}(t)\,w_{h,n} (x)
\,, \eqno(5.3)$$
yielding in the Fourier-Laplace domain
$$ \widehat{\widetilde {p}}_{h,\tau }(\kappa ,s) =
\sum_{n=0}^{\infty} \frac{1-\widetilde {\phi} (\tau s)}{s}\,
\lt( \widetilde {\phi} (\tau s)\rt)^n\,
\lt( \widehat {w} (h\kappa )\rt)^n\,.\eqno(5.4)$$
Separately we treat the powers $\lt( \widetilde \phi (\tau s)\rt)^n$
and   $\lt( \widehat w (h\kappa )\rt)^n$, so avoiding
the problematic simultaneous inversion of the diffusion limit
from the Fourier-Laplace domain into
the physical domain.

Observing    from (5.1)
$$    \lt( \widetilde \phi (\tau s)\rt)^n   \sim
    \lt( 1 - \lambda (\tau s)^\beta \rt)^n \,,\eqno(5.5) $$
we relate the running index $n$ to the presumed operational time
$t_*$ by
$$ n \sim \frac{t_* }{\lambda \,\tau ^\beta }\,, \eqno(5.6)    $$
and for \underbar{fixed} $s$ (as required by the continuity theorem
of probability theory), by sending
$\tau \to 0$ we get
$$ \lt( \widetilde \phi (\tau s)\rt)^n   \sim
\lt(1- \lambda \,\tau ^\beta s^\beta \rt)^{t_* /(\lambda \tau ^\beta )}
\to  \exp \lt(-t_*  \,s^\beta \rt)\,. \eqno(5.7)$$
Here $s$ corresponds to physical time $t$, and in Laplace inversion
we must treat $t_* $ as a parameter.
Hence, in physical time $\exp (-t_* s^\beta )$ corresponds
to
$$ \bar g_\beta(t, t_*) = t_* ^{-1/\beta } \, \bar g_\beta (t_* ^{-1/\beta } t)\,,
\eqno(5.8) $$
with $\widetilde{\bar g}_\beta (s) = \exp (-s^\beta )$.
Here $\bar g_\beta(t, t_*)$  is the totally positively skewed stable
   density (with respect to the variable $t$) evolving in operational
   time $t_*$ according to the "space"- fractional equation
$$\frac{\d}{\dt_*}\, \bar g_\beta(t, t_*) =
\,_tD^\beta_{-\beta} \,\bar g_\beta(t, t_*)\,,
\quad \bar g_\beta(t,0) = \delta(t)\,,
\eqno(5.9)$$
where $t$ is playing the role of the spatial variable.
Analogously, observing from (5.2)
$$    \lt( \widehat w (h\kappa )\rt)^n   \sim
 \lt( 1 - \mu  (h |\kappa| )^\alpha \, i ^{\ds \theta \sgn \kappa}\rt)^n \,,
\eqno(5.10)
$$
and with the aim of obtaining a meaningful limit
we now set
$$ n  \sim \frac{t_* }{\mu h^\alpha }\,, \eqno(5.11)$$
and find, by sending $h\to 0^+$, the relation
$$     \lt( \widehat w (h\kappa )\rt)^n
 \sim \lt(1-\mu (h |\kappa|)^\alpha\,i^{\ds \theta\sgn\kappa}\rt)
 ^{t_*/(\mu h^\alpha)}
 \to
   \exp \lt(-t_* |\kappa|^\alpha   \, i ^{\ds \theta \sgn \kappa} \rt),
    \eqno(5.12) $$
the Fourier transform of a $\theta$-skewed $\alpha$-stable
density $f_{\alpha,\theta} (x,t_* )$ evolving
in operational time $t_* $.
This density is the solution of the space-fractional equation
$$ \frac{\d}{\d t_* }\,f_{\alpha,\theta} (x,t_* )
   =  \,_xD_\theta^\alpha \, f_{\alpha,\theta} (x,t_* )\,,
 \quad  f_{\alpha,\theta} (x,0)=\delta (x)\,. \eqno(5.13) $$

The two relations (5.6) and (5.11)
between the running index $n$ and the presumed operational time $t_* $
require the (asymptotic) {\it scaling relation}
$$ \lambda \,\tau ^\beta \sim \mu \, h^\alpha\,,\eqno(5.14)  $$
that for purpose of computation we simplify to
$$ \lambda \,\tau ^\beta = \mu \, h^\alpha \,. \eqno(5.15)$$

Replacing $t_* $ by $t_{*,_n} = n \lambda \tau ^\beta$, using
the asymptotic results (5.7) and (5.12) obtained for the powers
$ \lt( \widetilde \phi (\tau s)\rt)^n  $ and
$ \lt( \widehat w (h\kappa )\rt)^n  $, furthermore noting
$$ \frac{1-\widetilde \phi (\tau s)}{s} \sim
  s^{\beta -1} \,\lambda \,\tau ^\beta \,, $$
we finally obtain from (5.4) the Riemann sum
(with increment $\lambda \tau^\beta$)
$$
 \widehat{\widetilde {p}}_{h,\tau }(\kappa ,s) \sim
s^{\beta -1} \,
\sum_{n=0}^\infty
\exp \lt[
 -n\lambda \tau ^\beta
\lt(s^\beta  +|\kappa |^\alpha i^{\ds \theta \sgn\kappa}\rt)\rt]
\,\lambda \,\tau ^\beta    \,, \eqno(5.16)
$$
and hence the integral
$$  \widehat{\widetilde {p}}_{h,\tau }(\kappa ,s) \sim
s^{\beta -1} \,
\int_{0}^\infty
\exp \lt[
 -t_* \lt(s^\beta  +|\kappa |^\alpha i^{\ds \theta \sgn\kappa}\rt)\rt]
\,dt_*\,. \eqno(5.17)$$
For the {\it limiting process} $u_\beta (x,t)$ this means
 $$
 \widehat{\widetilde {u}}_{\beta  }(\kappa ,s) =
\int_{0}^\infty s^{\beta -1} \,
\exp \lt[
 -t_* \lt(s^\beta  +|\kappa |^\alpha i^{\ds \theta \sgn\kappa}\rt)\rt]
\,dt_*\,. \eqno(5.18)$$
Observe that the RHS of this equation is just another way
of writing the RHS of equation (2.5) which is the Laplace-Fourier
solution of the STFD 
equation (2.1).
By inverting the transforms we get after some manipulations
(compare \cite{M3 PRE02sub}) in physical space-time
the {\it integral formula of  subordination}
$$
u_\beta (x,t)=
  \int_0^\infty
           f_{\alpha,\theta} (x,t_*)\, g_\beta(t_*,t)\, dt_*\, \eqno(5.19)$$
  with
$$ g_\beta(t_*,t) = \frac{t}{\beta}\,
 \bar {g}_\beta \lt(t\, t_* ^{-1/\beta}\rt) \,t_*^{-1/\beta -1} \eqno(5.20)  $$
  standing for the density $r_t (t_*)$ in equation (4.1).

There are two processes involved.
One is the unidirectional  motion along the
$t_*$  axis representing the operational time.
This motion happens in physical time $t$ and the $pdf$ for the
operational time having value $t_*$  is (as density in $t_*$,
evolving in physical time $t$) given by (5.20).
In fact, by substituting $y= t\, t_*^{-1/\beta }$  we find
 $$ \int_0^\infty  g_\beta (t_*,t)\, dt_* \equiv
 \int_0^\infty  \bar {g}_\beta (t,t_*)\, dt = 1\,, \quad  \forall \, t>0\,.
 \eqno(5.21) $$
The operational time $t_*$ stands in analogy to the
counting index $n$ in Eqs. (3.5) and (5.4).
The other process is the process described by Eq. (5.13), a spatial
probability density for sojourn of the particle
in point $x$ evolving in  operational time $t_*$,
$$ \bar u_\beta(x,t_*) = f_{\alpha,\theta} (x,t_*)\,.\eqno(5.22) $$
We get the solution to the Cauchy problem (2.1), namely the $pdf$
$u(x,t)=u_\beta (x,t)$
for sojourn in point $x$, evolving in physical time $t$, by {\it averaging}
$\bar u_\beta (x, t_*)$ with the {\it weight function}
$g_\beta(t_*,t)$ over the interval $0<t_*<\infty$ according to (5.19).

\section{Sample path for  space-time fractional diffusion}

In the series representation (3.5) of the CTRW the running index $n$
(the number of jumps having occurred up to physical time $t$) is a
{\it discrete operational time}, proceeding in unit steps. To this
index $n$ corresponds the physical time $t=t_n$, the sum of the first
$n$ waiting times, and in physical space the position
$x = x_n$, the sum of the first $n$ jumps, see Section 3.

Rescaling space and physical time by factor
$h$ and $\tau $, obeying the {\it scaling relation}
$$ \mu \,h^\alpha  = \lambda \, \tau ^\beta \,, \eqno(6.1)$$
and introducing, by sending $\{ h\to 0\,, \, \tau \to 0\}$,
{\it continuous operational time}
$$ t_*  \sim n\,\lambda \,\tau ^\beta \sim n\, \mu \, h^\alpha \,.
\eqno(6.2)$$
Then, in the series representation (3.5) we have 
{\it two discrete Markov processes} (discrete in operation time $n$),
namely a random walk in the space variable $x$, 
with jumps $X_n$, and another
random walk (only in positive direction)
of the physical time $t$, 
making a forward jump $T_n$ at every instant $n$.

In the diffusion limit the spatial process becomes an
$\alpha $-stable  process for the position
$x = \bar{x} =  \bar{x}(t_* )$, whereas the unilateral time
process becomes a unilateral (positively directed)
$\beta$-stable  process for the physical time
$ t = \bar{t} =   \bar{t}(t_* )$.
A sample path of a diffusing particle in physical coordinates can
be produced by combining in the $(t,x)$ plane the two random functions
$$
\cases{
x = \bar{x} =  \bar{x}(t_* ) \,, \cr
t = \bar{t} =  \bar{t}(t_* ) \,, \cr
}
\eqno(6.3)
$$
both evolving in operational time $t_* $, both being Markovian and
obeying stochastic differential equations
$$
\cases{
d \bar{x} =  d (\hbox{L\'evy  noise of order} \, \alpha \;
     \hbox{and skewness}\, \theta)  \,, \cr
d \bar{t} =  d (\hbox{one sided L\'evy  noise of order} \, \beta)   \,. \cr
} \eqno(6.4) $$
This gives us in the $(t, x)$ plane
the $t_*\,$- parametrized particle path, and by elimination of
        $t_*$    we get it as $x=x(t)$.
We suggest to call this procedure "construction of a   particle path by 
{\it parametric subordination}".
  Note that the process  $t = T(t_*)$ yielding the second
   random function in (6.3)  has the properties of
   a {\it subordinator} in the sense of Definition 21.4 in
           \cite{Sato 99}.
\\
\underbar{Concerning notation}: It is good to make a conceptual
distinction between the position
$\bar{x}$ of an individual particle and the variable $x$,
likewise between the  physical time   position
$\bar{t}$ and the physical time variable $t$.
When there are many particles  we have overall densities for them
and for these densities fractional diffusion equations.
The $pdf$ for the particle being in point $\bar{x} = x$
at operational time $t_* $, that we denote by
 $\bar{u}_\beta (x, t_* )= f_{\alpha,\theta} (x,t_*)$,
satisfies the evolution equation
(Eq. (5.13) re-written with  $\bar{u}_\beta$)
$$
\frac{\d}{\d t_* }\,\bar{u}_\beta(x,t_* ) =
\,_xD_\theta^\alpha \, \bar{u}_\beta (x,t_* )
  \,, \quad \bar{u}(x,0) =\delta (x)\,.
\eqno(6.5)$$
The $pdf$ for the  physical time being in  $\bar{t} = t$
at operational time $t_* $, that we denote by
$\bar{v}(t,t_* )= \bar g_\beta (t,t_*)$,
 obeys the skewed fractional equation
$$
\frac{\d}{\d t_* }\,\bar{v}(t,t_* ) =
   \,_t D_{-\beta} ^\beta \, \bar{v}(t,t_* )\,,
\quad \bar{v} (t,0) = \delta (t)\,.
\eqno(6.6)$$
\underbar{Remind}:
In operational time two Markovian random functions
$\bar{x}(t_* )$, $ \bar{t}(t_* )$ occur, as random processes,
individually
for each particle. In physical coordinates we have
the $t_* $-parametrized random path described by (6.3).
\\
\underbar{Remark}:
It is instructive to see what happens for the limiting value
      $\beta = 1$. In this case the Laplace transform of
     $\bar{g}_\beta(t,t_*) = \bar{g}_1 (t,t_*)$ is $\exp(-t_* s)$,
      implying $ \bar{g}_1 (t,t_*) = \delta(t-t_*)$, the delta density
      concentrated on $t=t_*$. So, in this case, $t=t_*$, operational
      time and physical time coincide.
\section{Numerical results}

In this  Section,
after describing the numerical schemes adopted,
we shall show the sample paths for four case studies
of symmetric ($\theta=0$) fractional diffusion processes:
$\{\alpha =2,  \, \beta =0.90\}$,
$\{\alpha =2,  \, \beta =0.80\}$,
$\{\alpha =1.5, \,  \beta =0.90\}$,
$\{\alpha =1.5, \,  \beta =0.80\}$.
As explained in the previous Sections, for each case
we need to construct the sample paths for three  distinct processes,
the parent process $x= Y(t_*)$, the leading process
$t = T(t_*)$ (both in the operational time)
and, finally, the subordinated process $x =X(t)$,
corresponding to the required fractional diffusion process.
For this purpose we  proceed as follows 
for the required three steps.

First, let  the operational time $t_*$ assume
 $N$ discrete equidistant values in a given interval $[0,T]$,
that is
$t_{*,n}=n T/N,\;  n=0, 1,\dots,N$.
As a working choice we take $T=1$ and $N=10^{6}$.
Then produce $N$ independent
identically distributed ($iid$) random deviates,
 $Y_1,Y_2,\dots,Y_{N}$ having a symmetric
stable probability  distribution of order $\alpha$,
see the book by Janicki \cite{Janicki LN96}
for a useful and efficient method to do that.
Now, with the  points
 $$x_0=0,\quad  x_n = \sum\limits_{k=1}^n X_k,\quad n=1,\dots,N\,,
\eqno(7.1)$$
the couples $(t_{*,n},x_n)$, plotted
in the  $(t_*,x)$ plane  (operational time, physical space)
can be considered as points of
a true sample path
$\{x(t_*) : 0\le t_*\le T\}$
of a symmetric L\'evy motion with order  $\alpha$
corresponding
to the integer values of operational time $t_*=t_{*,n}$.
In this identification of $t_*$ with $n$ we use the fact that
our stable laws for waiting times and jumps imply
$\lambda = \mu = 1$  in the asymptotics (5.1) and (5.2) and
$\tau = h = 1$ as initial scaling factors in (5.3) and (5.14).

In order to complete the sample path we agree to connect every
two successive  points
$(t_{*,n},x_n)$  and  $(t_{*,n+1},x_{n+1})$
by a horizontal line from $(t_{*,n},x_n)$ to
$(t_{*,n+1},x_n),$ and a vertical line from
$(t_{*,n+1},x_n)$ to $(t_{*,n+1},x_{n+1}).$
Obviously, this is not the 'true' L\'evy motion from point
$(t_{*,n},x_n)$ to point $(t_{*,n+1},x_{n+1})$, but from the
theory of CTRW we know this kind of discrete random process
to converge in the appropriate sense to  L\'evy motion.
The points $(t_{*,n}, x_n)$ are points of  a true L\'evy motion.

As  a second step, we  produce $N$ $iid$  random deviates,
  $T_1,T_2,\dots,T_{N}$ having a stable probability
distribution with order $\beta$ and skewness $ -\beta$
(extremal stable distributions).
Then, consider the points
  $$t_0=0,\quad  t_n = \sum\limits_{k=1}^n T_k,\quad n= 1,\dots,N\,,
\eqno(7.2)
$$
and plot the couples $(t_{*,n},t_n)$
in the  $(t_*,t)$ (operational time, physical time) plane.
By  connecting  points with
horizontal and vertical lines we get   sample paths
$\{t(t_*) : 0\le t_*\le N \tau = 1 \}$
describing the evolution of the physical time $t$ with
increasing operational time $t_*.$

The final (third) step consists in
plotting  points $(t(t_{*,n}),x(t_{*,n}))$ in the  $(t,x)$
plane, namely the physical time-space plane, and connecting
them as before.
So one gets a good   approximation of
the sample  paths of the subordinated fractional diffusion
process of parameters $\alpha$, $\beta$ and $\theta=0$.



Now as the successive values of $t_{*,n}$ and $x_n$ are
generated by successively adding the relevant standardized
stable random deviates,  the obtained sets of points in the
three coordinate planes:
$(t_*, t)$, $(t_*,x)$, $(t,x)$
can, in view of infinite divisibility and self-similarity of the stable
probability distributions, be considered as snapshots of
the corresponding true random processes occurring in
continuous operational time $t_*$ and physical time $t$, correspondingly.
Clearly, fine details between successive points are missing.
They are hidden:
\\
- In the $(t_*,x)$ plane in the horizontal lines from
$(t_{*,n},x_n)$ to $(t_{*,n+1},x_n)$ and the vertical
lines from $(t_{*,n+1},x_n)$ to $(t_{*,n+1},x_{n+1})$.
\\
- In the $(t_*,t)$ plane in the horizontal lines from
$(t_{*,n},t_n)$ to $(t_{*,n+1},t_n)$ and the vertical
lines from $(t_{*,n+1},t_n)$ to $(t_{*,n+1},t_{n+1})$.
\\
- In the $(t,x)$ plane in the horizontal lines from
$(t_n,x_n)$ to $(t_{n+1},x_n)$ and the vertical lines
from $(t_{n+1},x_n)$ to $(t_{n+1},x_{n+1})$.

The well-scaled passage to
the diffusion limit here consists simply in regularly subdividing the
$\{t_*\}$ intervals  of length 1   into smaller and smaller
subintervals (all of equal length   $\tau$
and adjusting the random
increments of $t$ and $x$ according to the requirement of
self-similarity, namely taking, respectively,
the  waiting times and spatial jumps as $\tau^{1/\beta}$
multiplied by a standard extreme $\beta$-stable deviate,
$\tau^{1/\alpha}$ multiplied by a standard (in our special
case: symmetric) $\alpha$-stable deviate, respectively, as
required by the self-similarity properties of the stable
probability distributions).
Furthermore if we watch sample path in a large interval of
operational time $t_*,$ the points $(t_{*,n},x_n)$ and
$(t_{*,n+1},x_{n+1})$ will in the graphs appear very near to each
other  in operational time $t_*$ and aside from missing mutually
cancelling jumps up and down (extremely near to each other) we
have a good picture of the true processes.

\begin{figure}
\begin{center}
 \includegraphics[width=.52\textwidth]{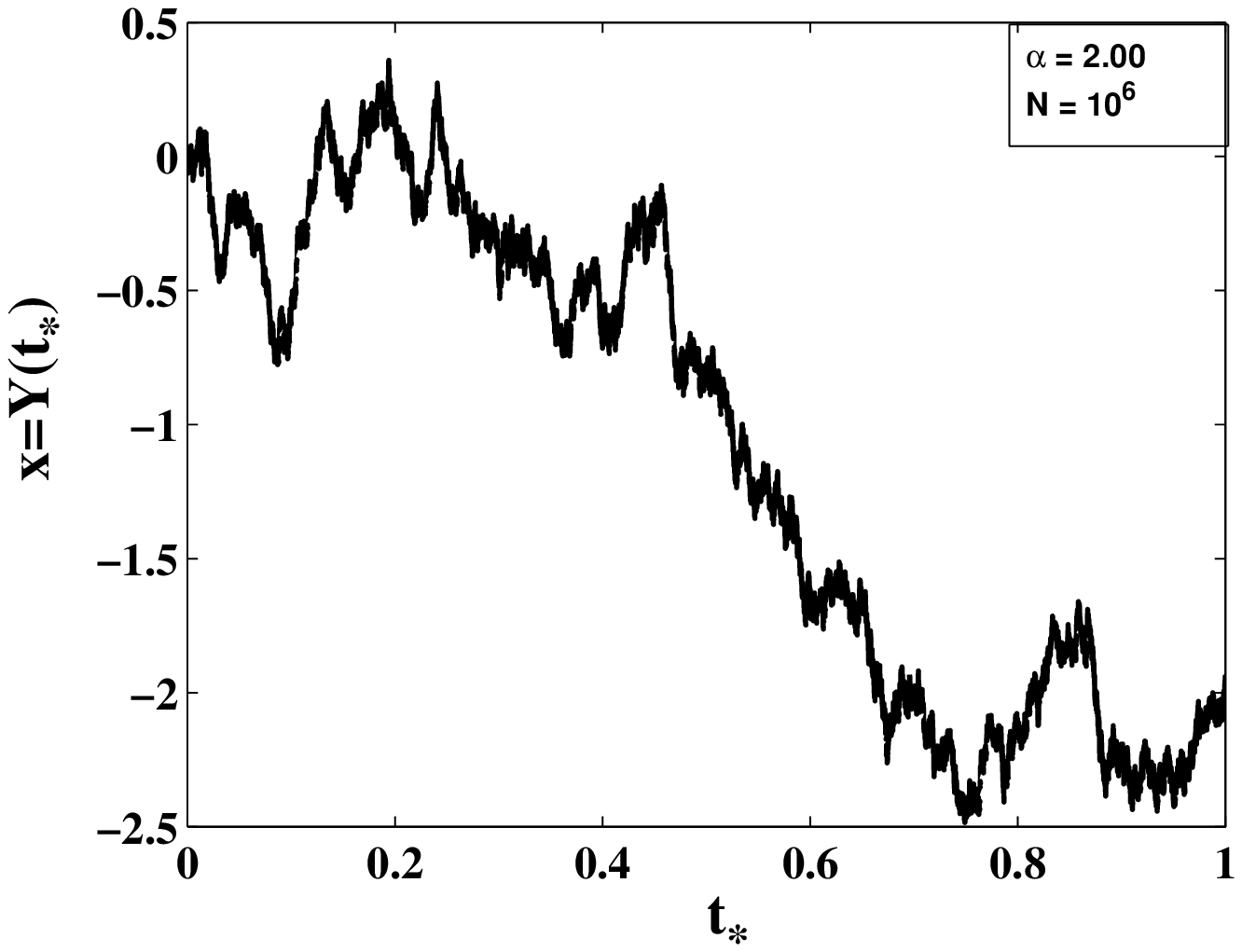}
\end{center}
 \caption{A sample path for the parent process $x=Y(t_*)$ with
 $\{\alpha =2\}$.}

\vskip 0.30truecm
 \includegraphics[width=.52\textwidth]{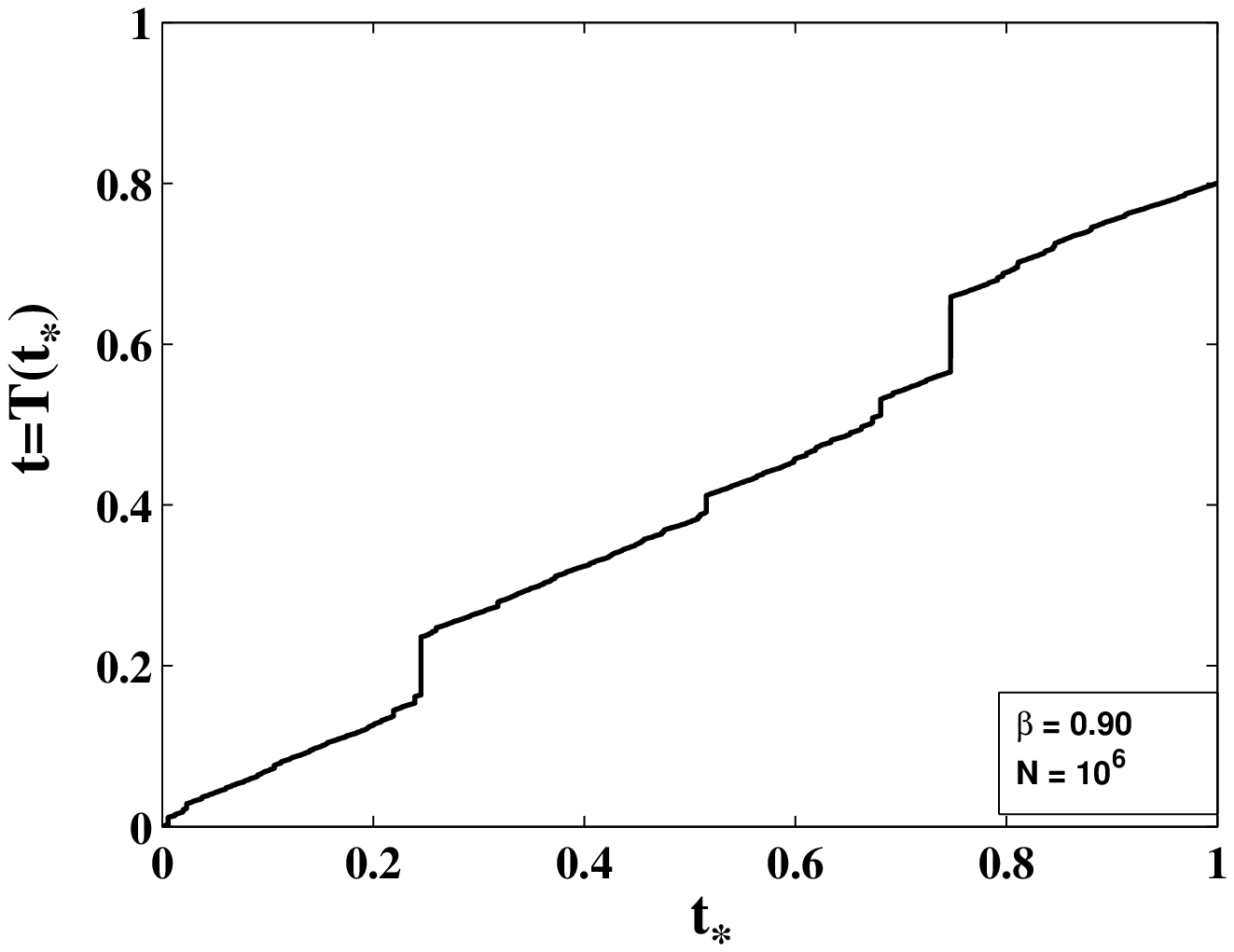}
\includegraphics[width=.52\textwidth]{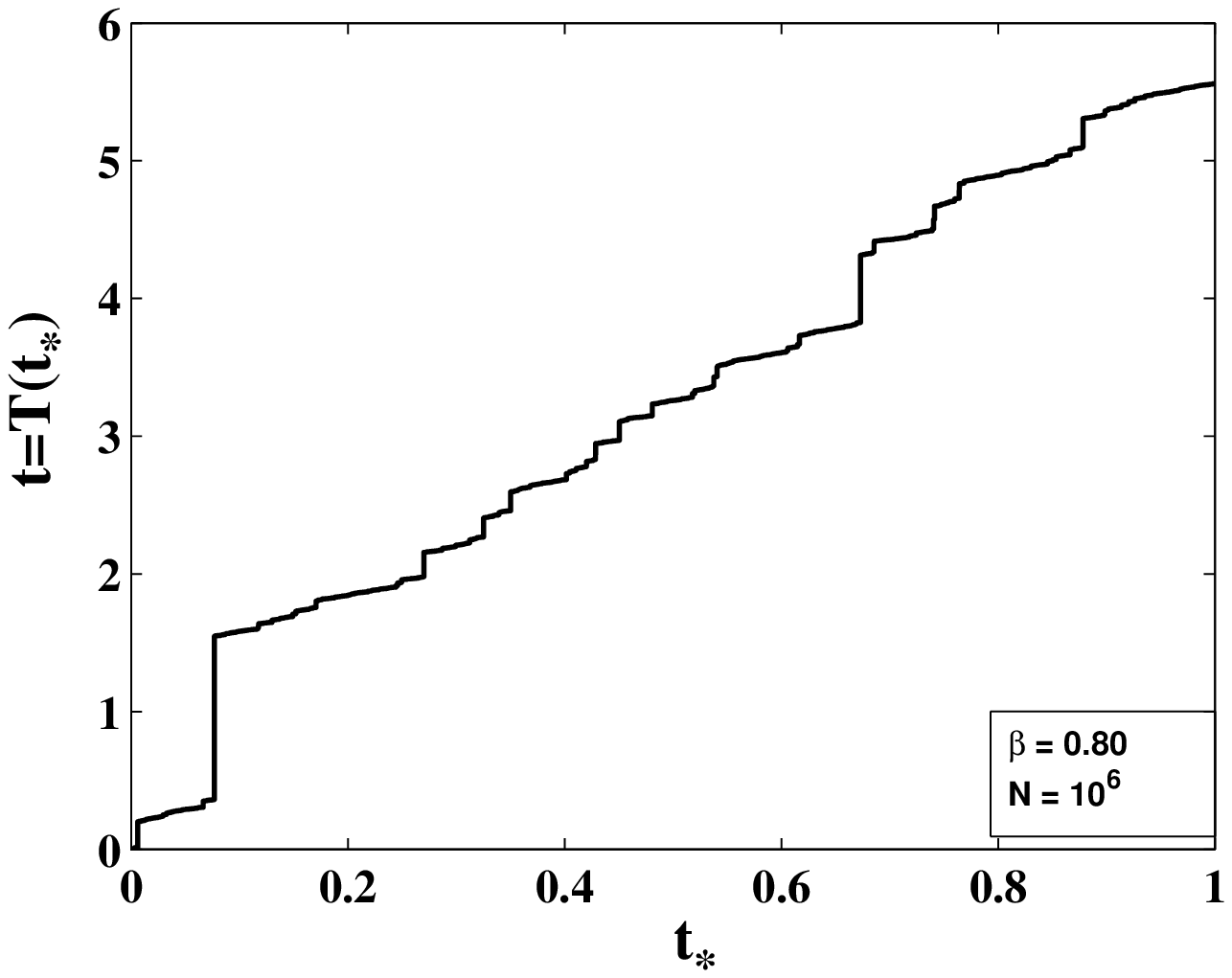}
 \caption{A sample path for the leading process $t=T(t_*)$.}
 \centerline{LEFT: $\{\beta =0.90 \}$,
      RIGHT: $\{\beta =0.80 \}$.}

\vskip 0.30truecm
 \includegraphics[width=.52\textwidth]{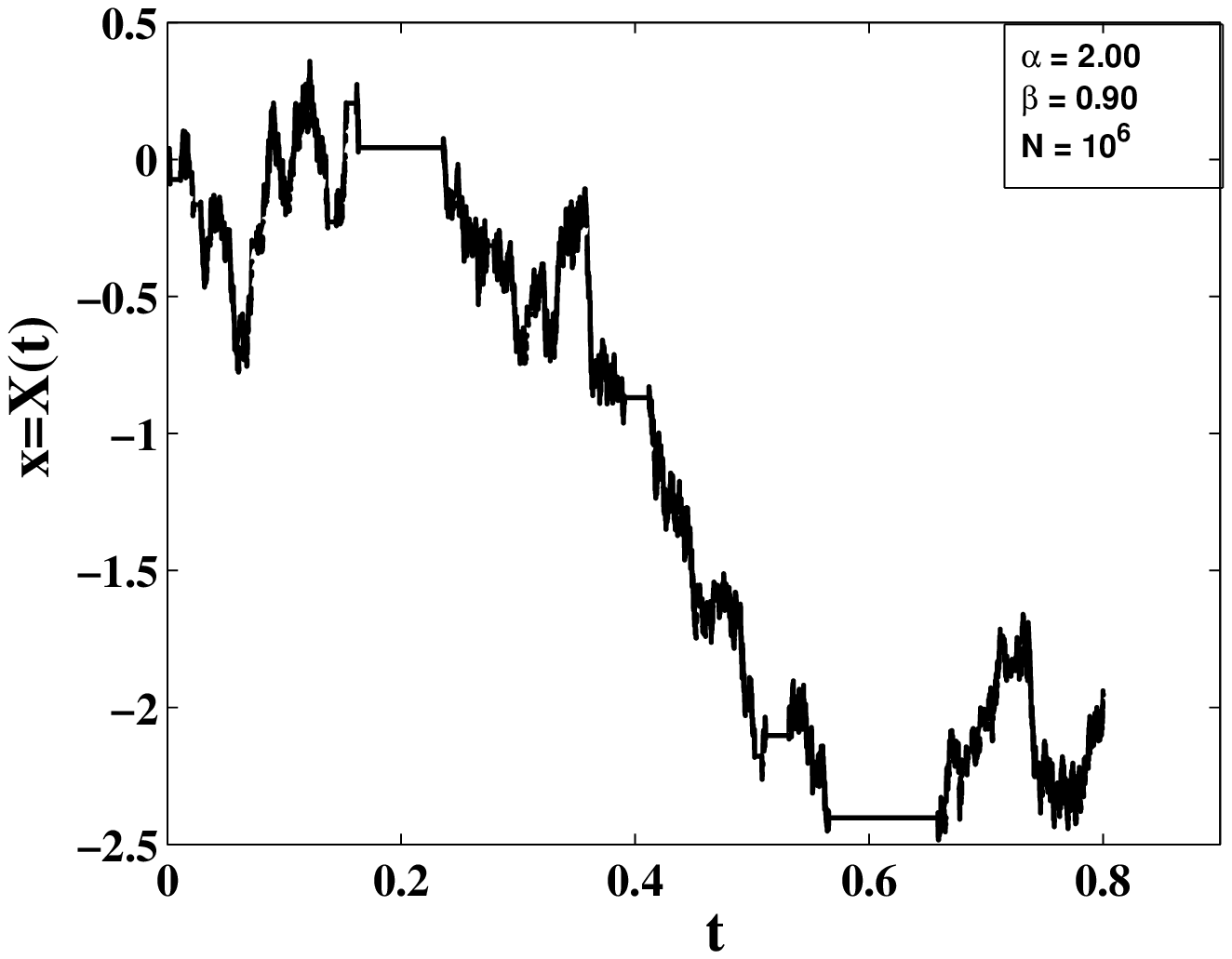}
\includegraphics[width=.52\textwidth]{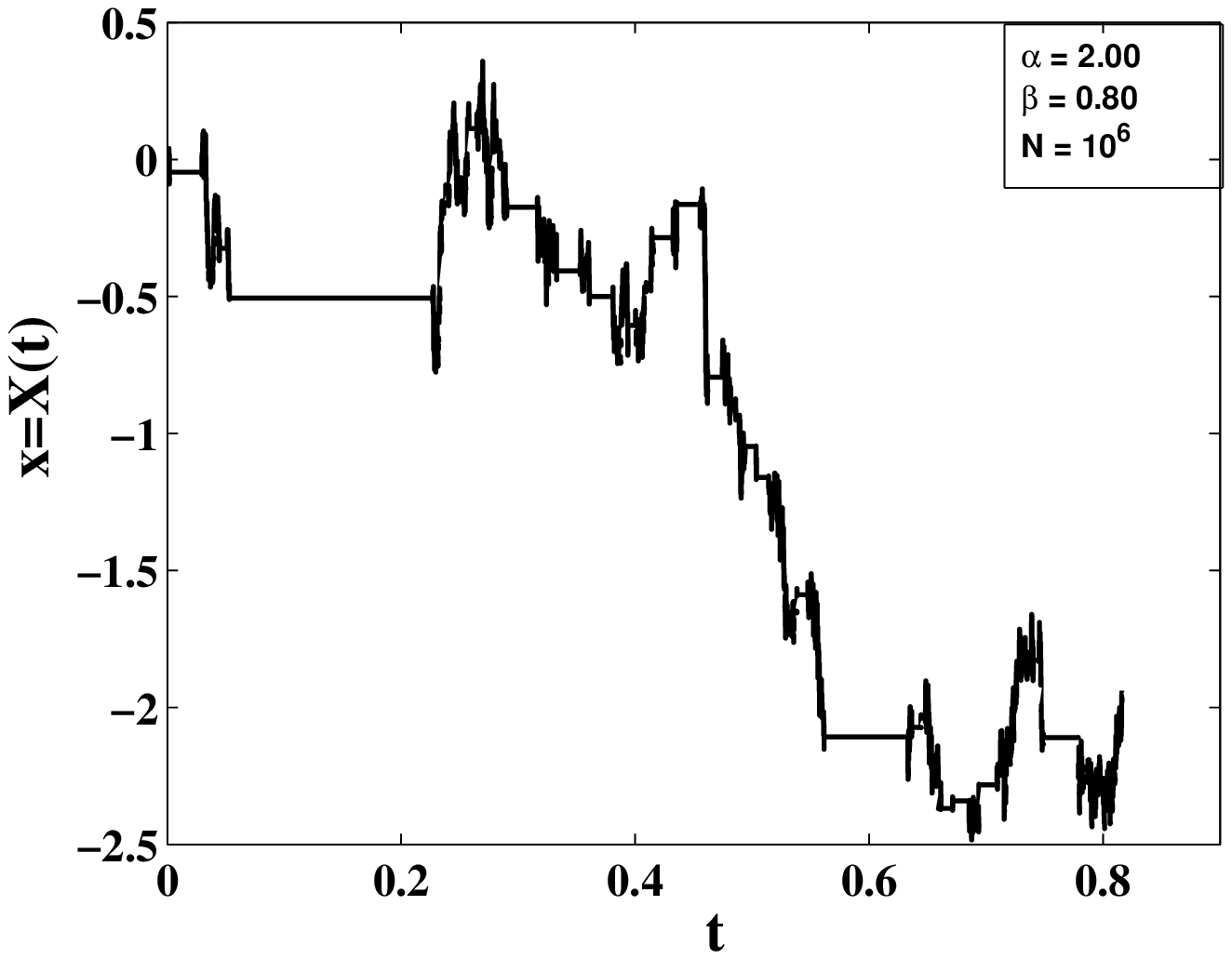}
 \caption{A sample path for the subordinated process $x=X(t)$.}
 \centerline{LEFT: $\{\alpha =2\,,\; \beta =0.90 \}$,
      RIGHT: $\{\alpha =2\,,\; \beta =0.80 \}$.}
\end{figure}


\begin{figure}
\begin{center}
 \includegraphics[width=.52\textwidth]{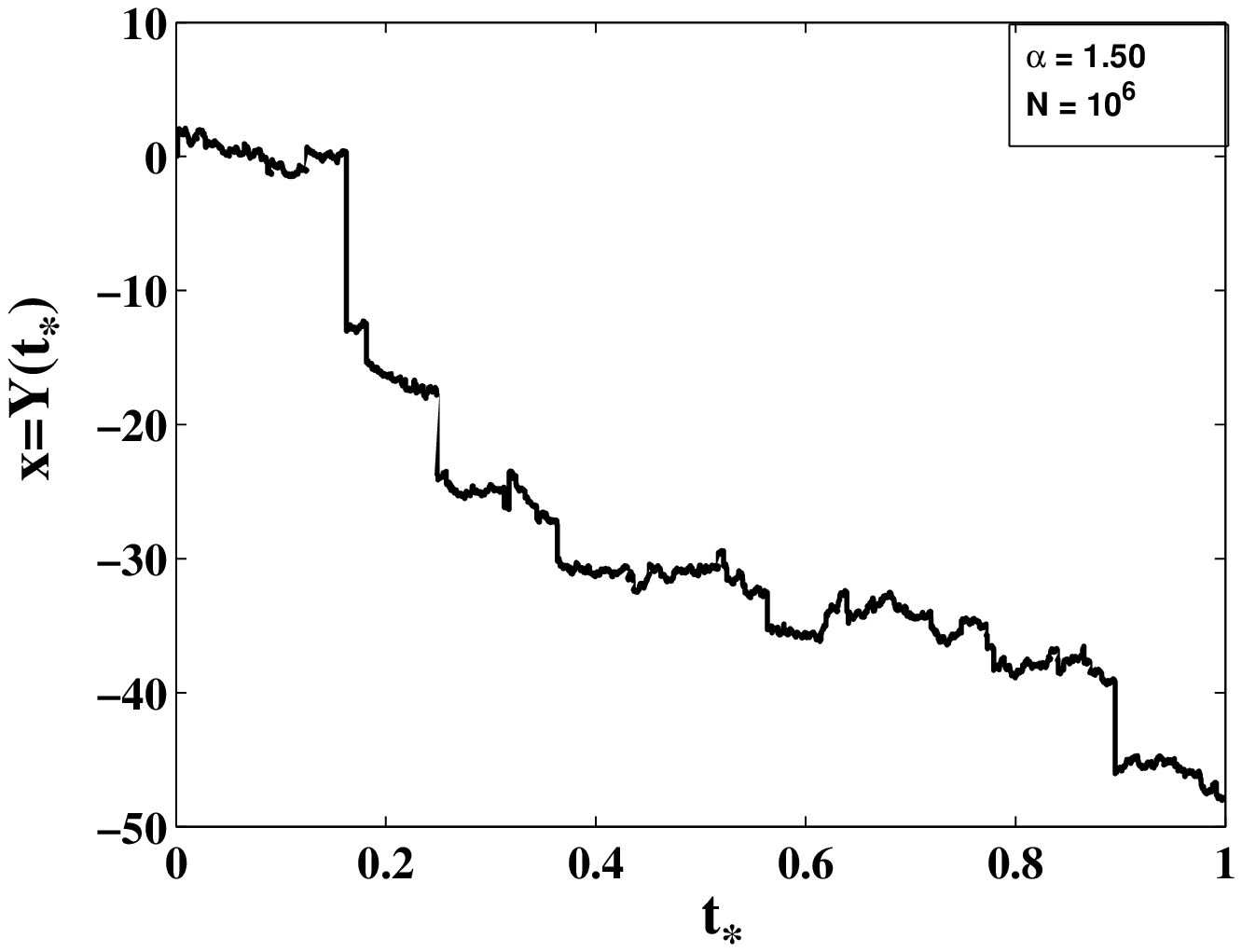}
\end{center}
 \caption{A sample path for the parent process $x=Y(t_*)$ with
 $\{\alpha =1.5\}$.}

\vskip 0.30truecm
 \includegraphics[width=.52\textwidth]{dp_90_6.eps}
\includegraphics[width=.52\textwidth]{dp_80_6.eps}
 \caption{A sample path for the leading process $t=T(t_*)$.}
 \centerline{LEFT: $\{\beta =0.90 \}$,
      RIGHT: $\{\beta =0.80 \}$.}

\vskip 0.30truecm
 \includegraphics[width=.52\textwidth]{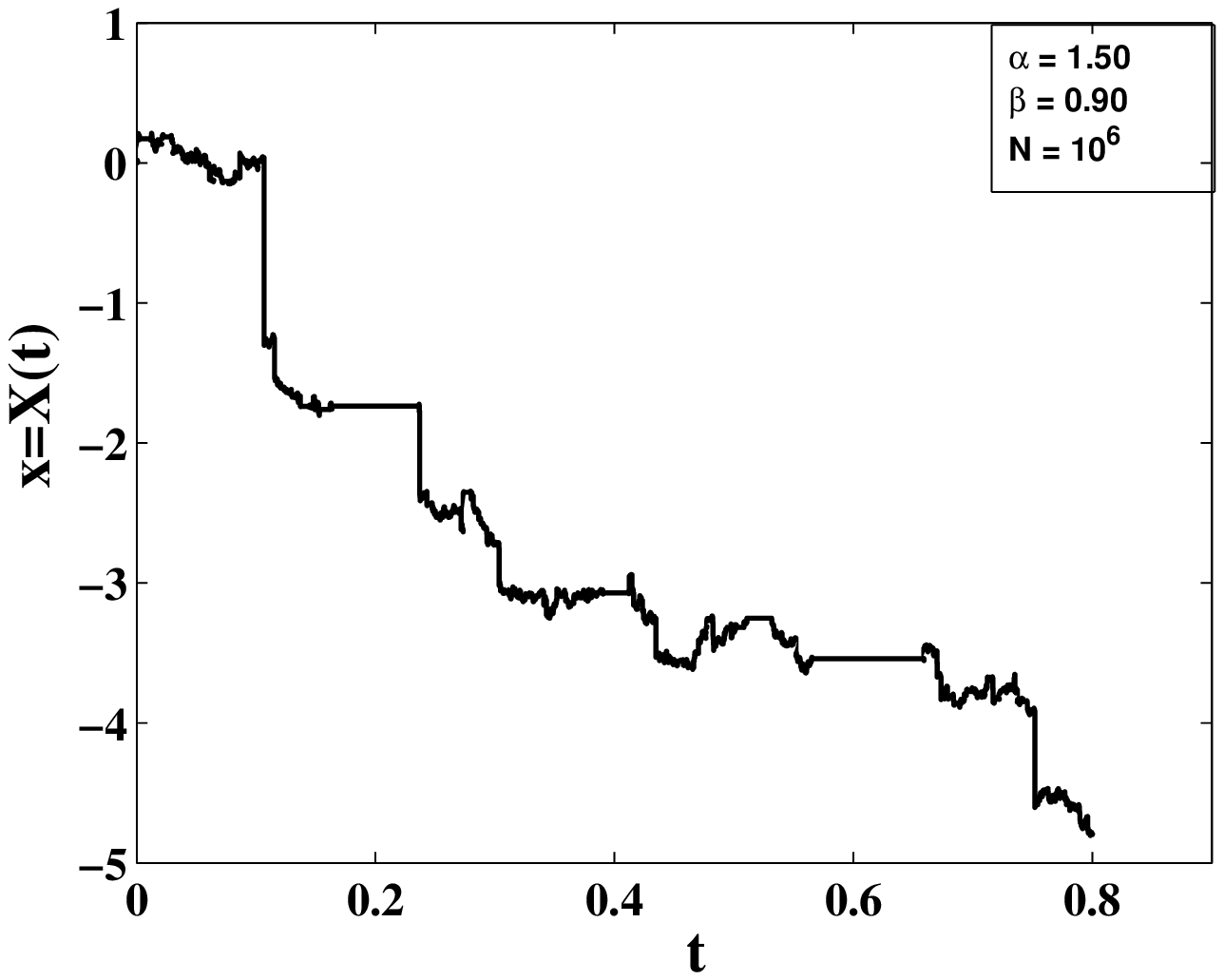}
\includegraphics[width=.52\textwidth]{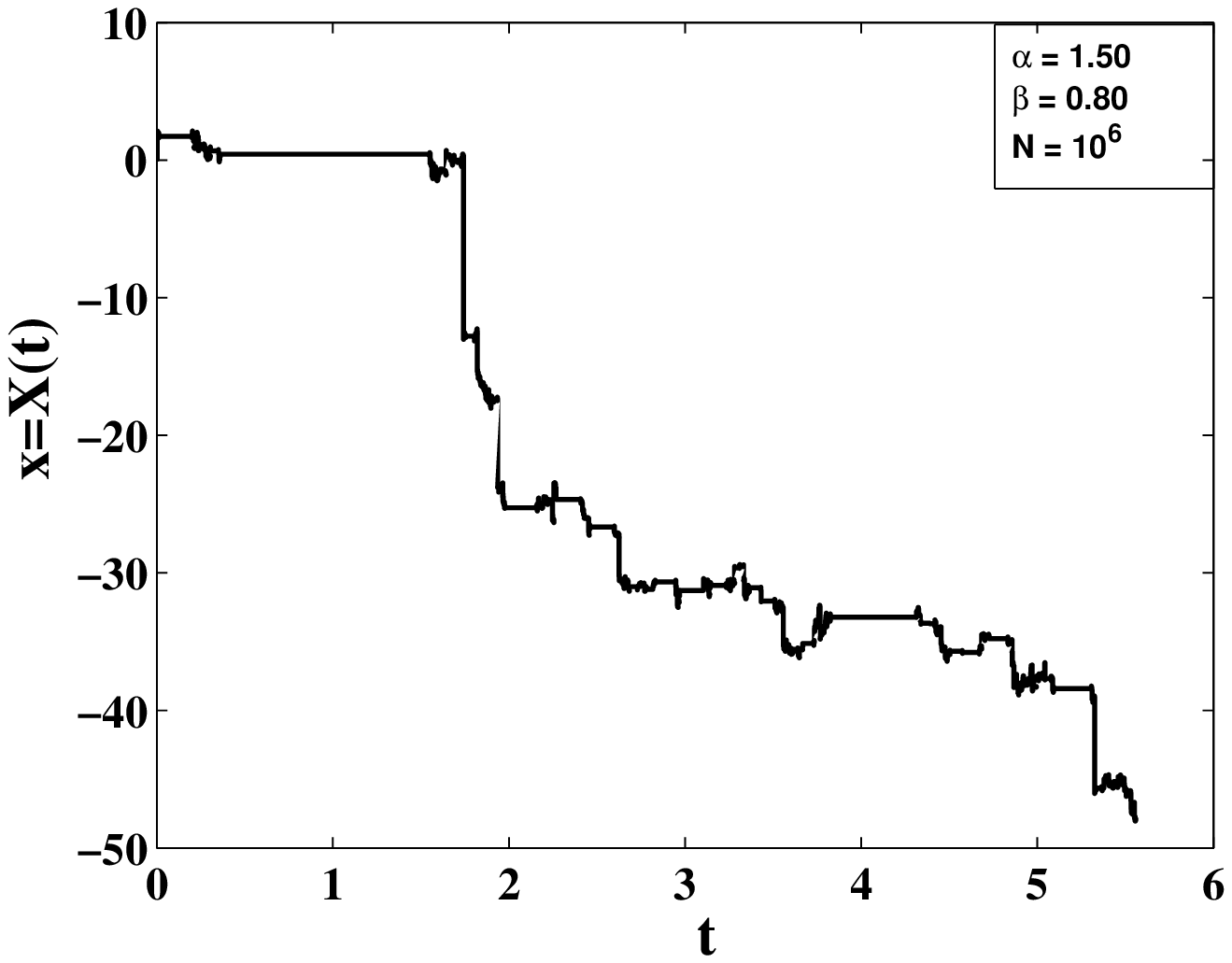}
 \caption{A sample path for the subordinated process $x=X(t)$.}
 \centerline{LEFT: $\{\alpha =1.5,\; \beta =0.90 \}$,
      RIGHT: $\{\alpha =1.5,\; \beta =0.80 \}$.}
\end{figure}

The resulting sample paths for all the processes involved in
the  two case studies are presented in the  Figs. 1-6.
The figure captions should clarify our strategy.
Figs. 1 and 4 are referring to the parent processes characterized
by the parameter  $\alpha =2$ and $\alpha =1.5$.
Figs. 2 and 5   are devoted to the leading processes\footnote{
Figs. 2 and 5 alternatively can also be viewed as graphical representations
of the {\it directing} processes $t_*=T_*(t)$ in the sense of Feller,
see  Section 4.
We note that the {\it directing} processes,
exhibiting horizontal segments, are no longer L\'evy processes even if
the random functions $t_*=T_*(t)$ are  non-decreasing and
right-continuous like the {\it leading}  processes $t=T(T_*)$.
This explains the non-Markovianity of the
subordinated processes.}
characterized by the parameter  $\beta  =0.9$ and $\beta  =0.8$
in the Right and Left plates, respectively.
As a consequence Figures 2 and 5 are identical, because are referring
to the same processes.
Finally,  Figs. 3 and 6 are devoted to the subordinated processes
resulting from the previous parent and leading processes.
Specifically in Fig. 3 the Left and Right plates show sample
paths for $\alpha =2$ and $\beta =0.9, 0.8$, respectively, and in Fig. 6
the Left and Right plates show sample paths for $\alpha =1.5 $
and $\beta =0.9, 0.8$, respectively.

By observing the figures the reader will note that horizontal
segments (waiting times) in the $(t, x)$ plane (Fig. 3, Fig. 6)
correspond to vertical segments (jumps) in the $(t, t_*)$ plane
(Fig. 2, Fig.5).
Actually, the graphs in the $(t, x)$-plane depict continuous time
random walks with waiting times $T_k$ (shown as horizontal segments)
 and jumps $X_k$ (shown as vertical segments). The left endpoints of
the horizontal segments can be considered as snapshots of the true
particle path (the true random process to be simulated),
the segments being segments of our ignorance.
In the interval $t_n < t \le t_{n+1}$ the true process
(namely the spatial variable $x=X(t)$) may jump up and
down (infinitely) often, the sum (or integral) of all
these ups (counted positive) and downs (counted negative)
amounting to the vertical jump $X_{n+1}$.

Finer details will become visible by choosing in the operational
time $t_*$ the step length $\tau$ smaller and smaller.
In the graphs we can clearly see what happens for finer and finer
discretization of the operational time $t_*$, by adopting
$10^1$, $10^2$, $10^3$ steps, see Figures 7-18.
As a matter of fact  there is no visible difference in
the transition  for the successive decades $10^4$, $10^5$, $10^6$
steps as the great majority of spatial jumps and waiting
times are very small. This property also explains the visible
persistence of large jumps and waiting times even of very
small steps $\tau$ of the operational time.
\begin{figure}
 \includegraphics[width=.52\textwidth]{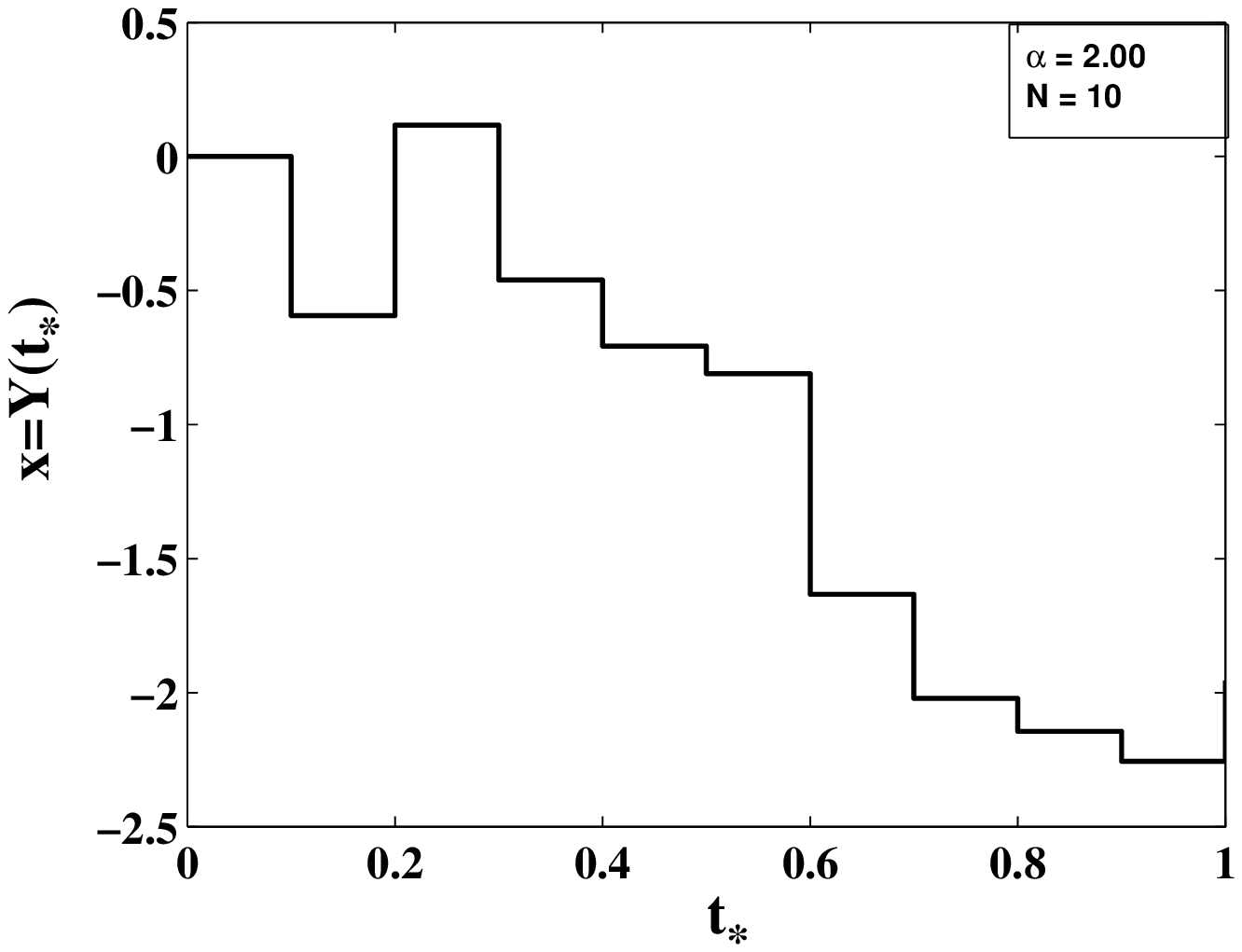}
\includegraphics[width=.52\textwidth]{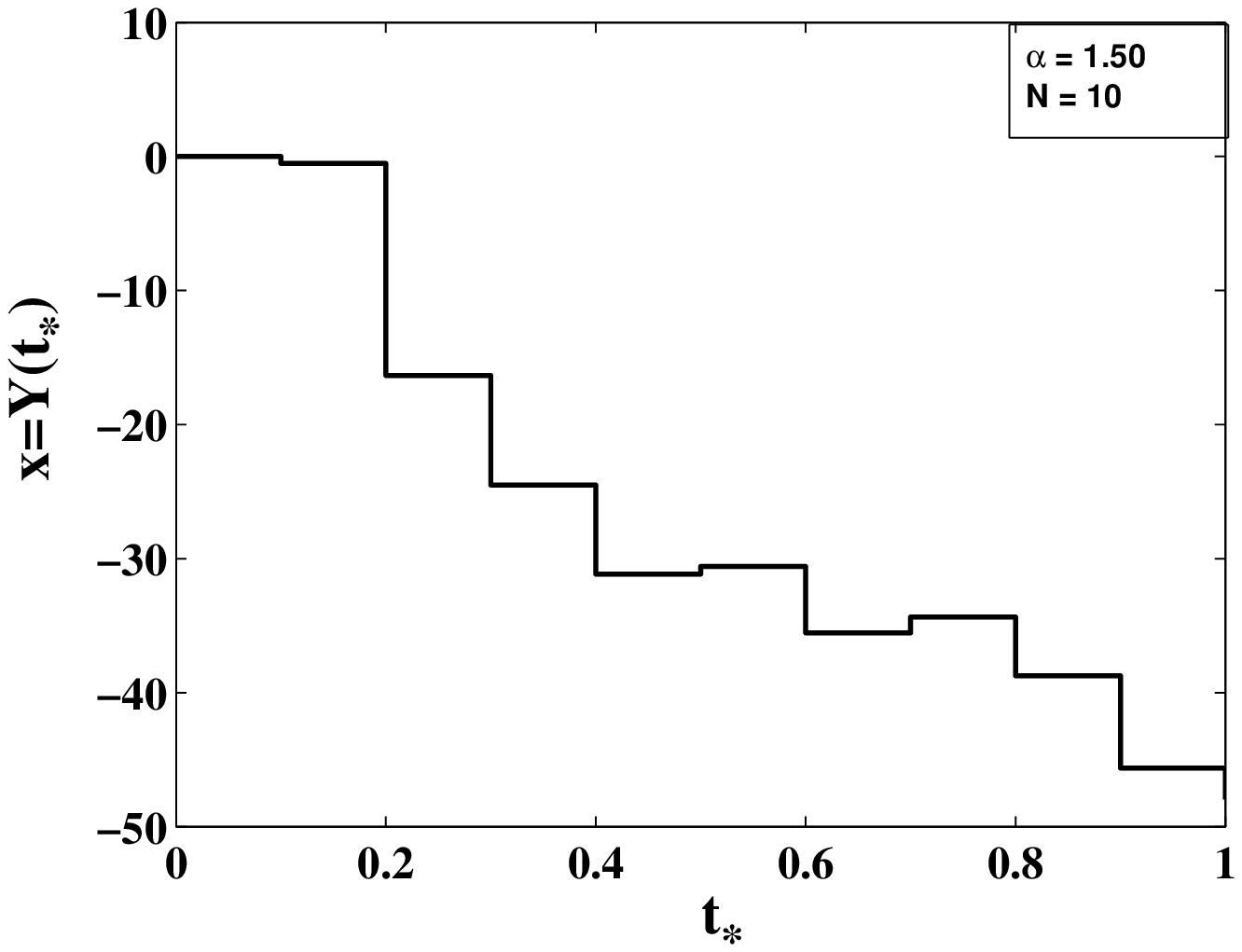}
 \caption{A sample path for the parent process $x=Y(t_*)$.}
 \centerline{LEFT: $\{\alpha =2,\; N = 10^1 \}$,
      RIGHT: $\{\alpha =1.5,\; N= 10^1 \}$.}

\vskip 0.30truecm
 \includegraphics[width=.52\textwidth]{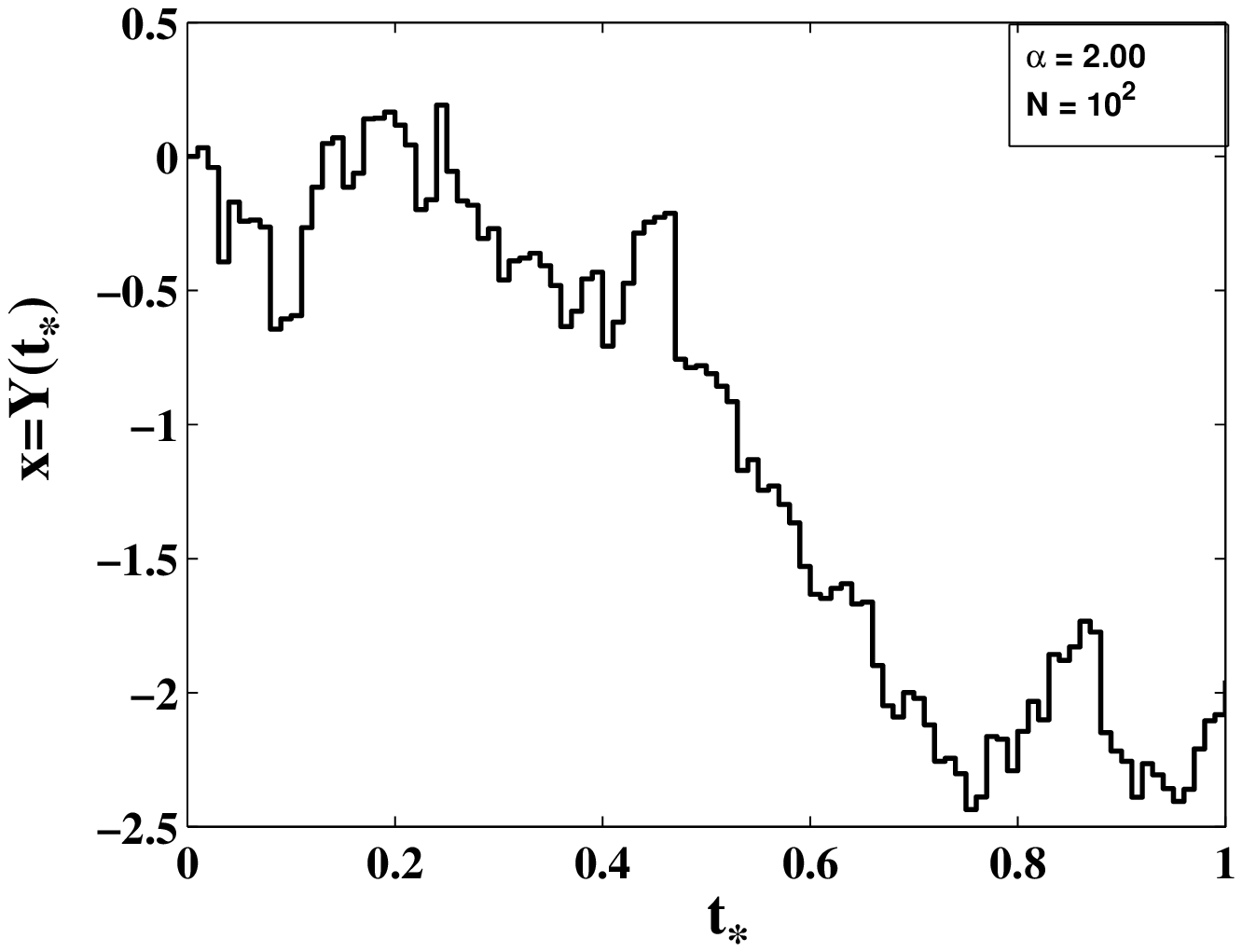}
\includegraphics[width=.52\textwidth]{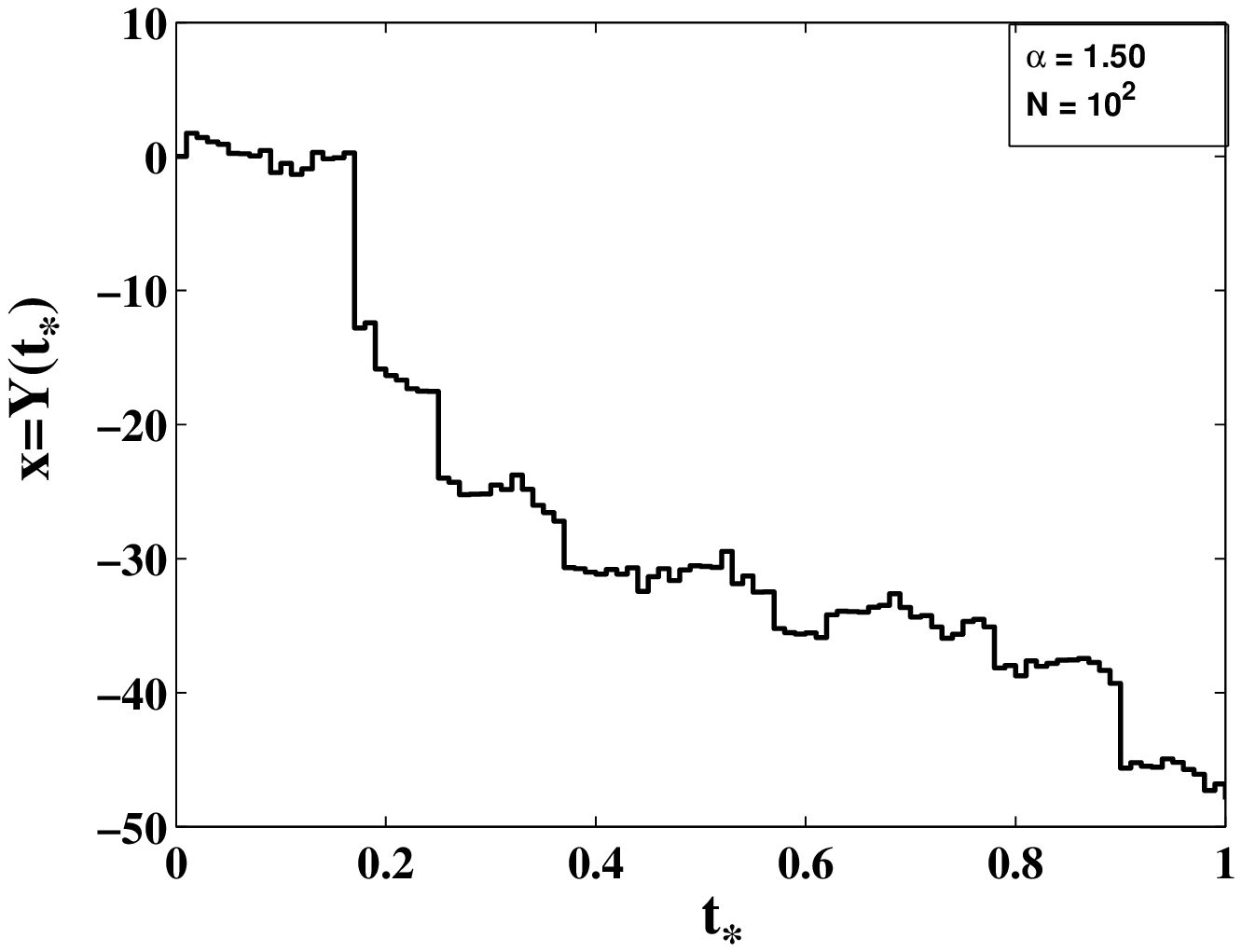}
 \caption{A sample path for the parent process $x=Y(t_*)$.}
 \centerline{LEFT: $\{\alpha =2,\; N = 10^2 \}$,
      RIGHT: $\{\alpha =1.5,\; N= 10^2 \}$.}

\vskip 0.30truecm
 \includegraphics[width=.52\textwidth]{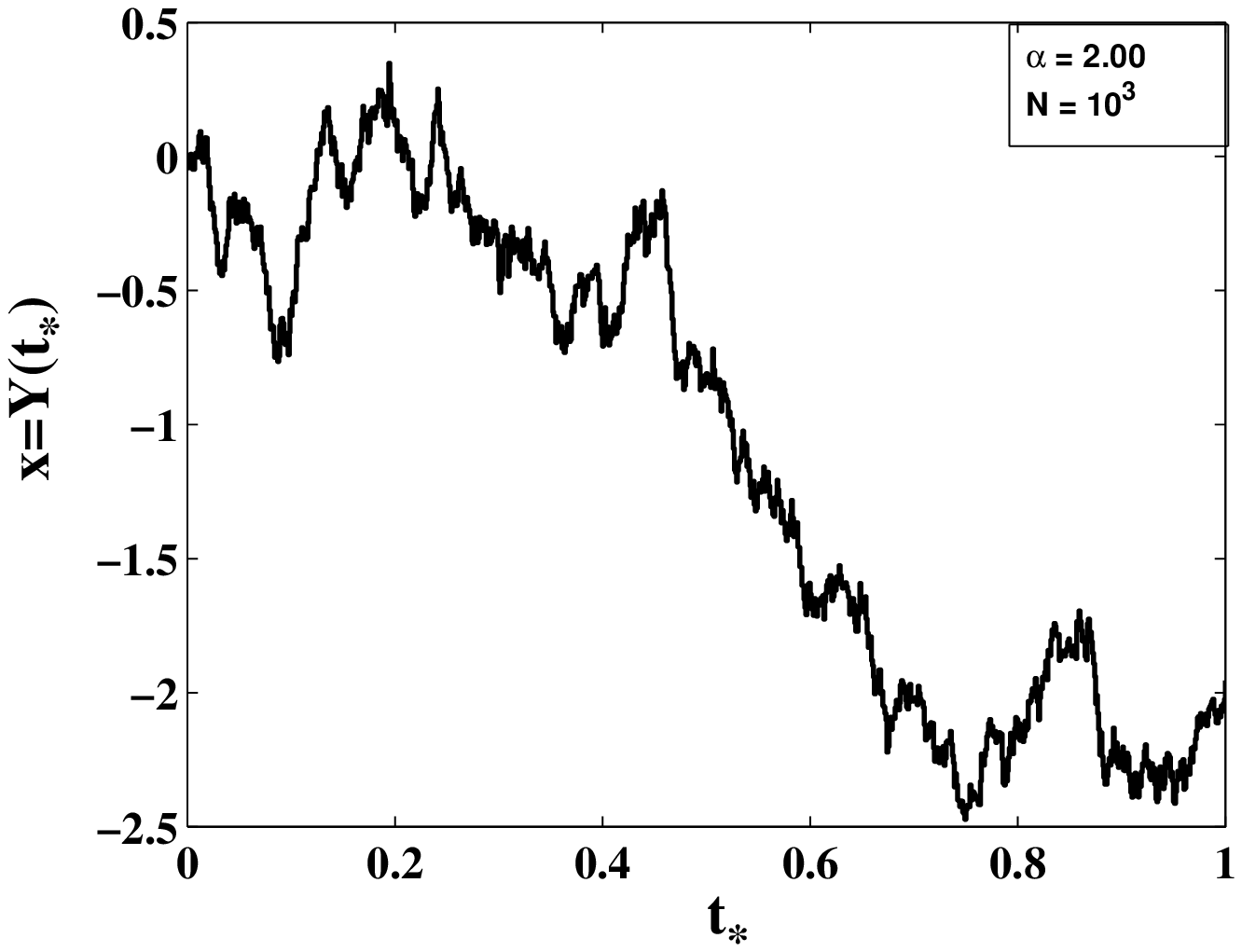}
\includegraphics[width=.52\textwidth]{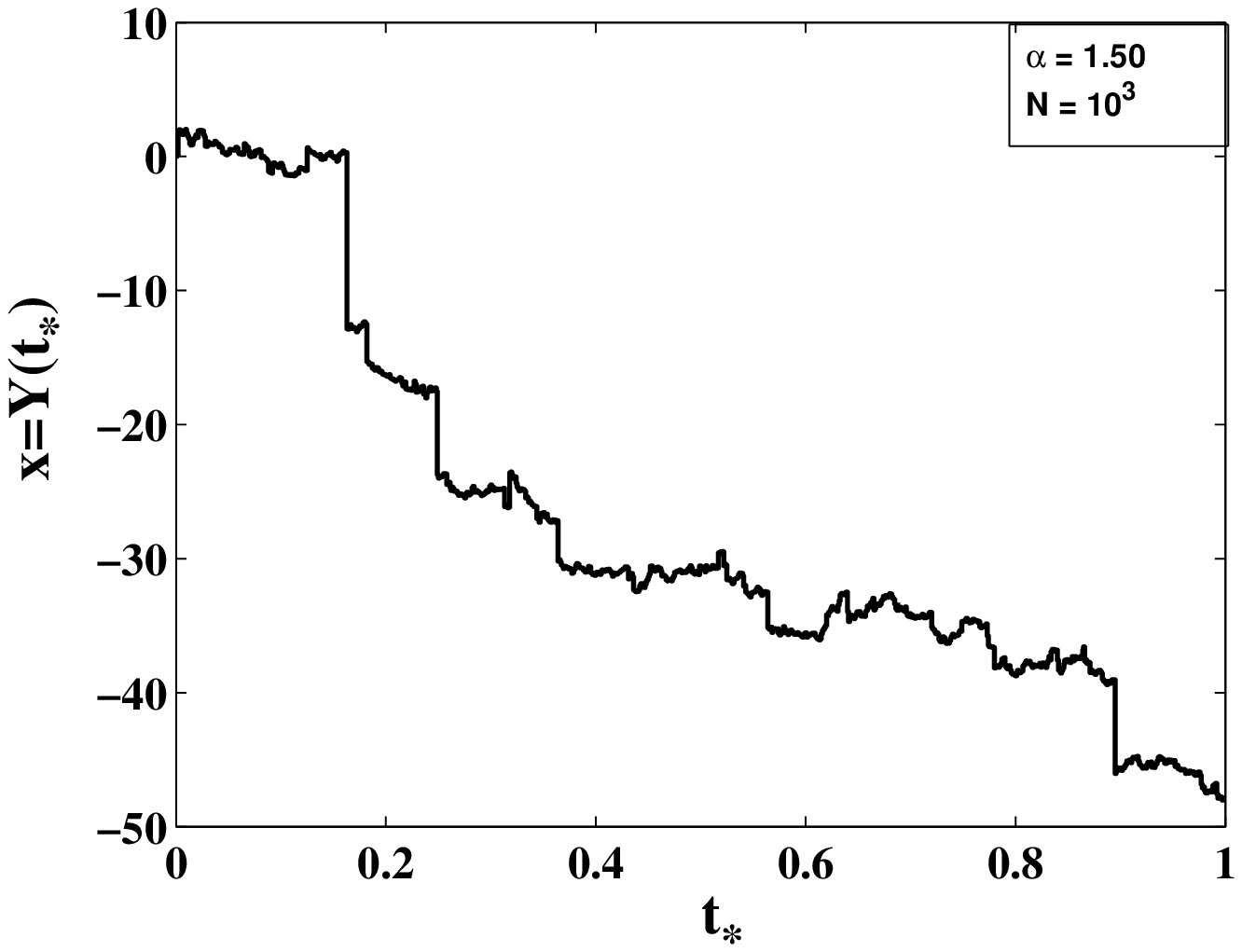}
 \caption{A sample path for the parent process $x=Y(t_*)$.}
 \centerline{LEFT: $\{\alpha =2,\; N = 10^3 \}$,
      RIGHT: $\{\alpha =1.5,\; N= 10^3 \}$.}
\end{figure}


\begin{figure}
 \includegraphics[width=.52\textwidth]{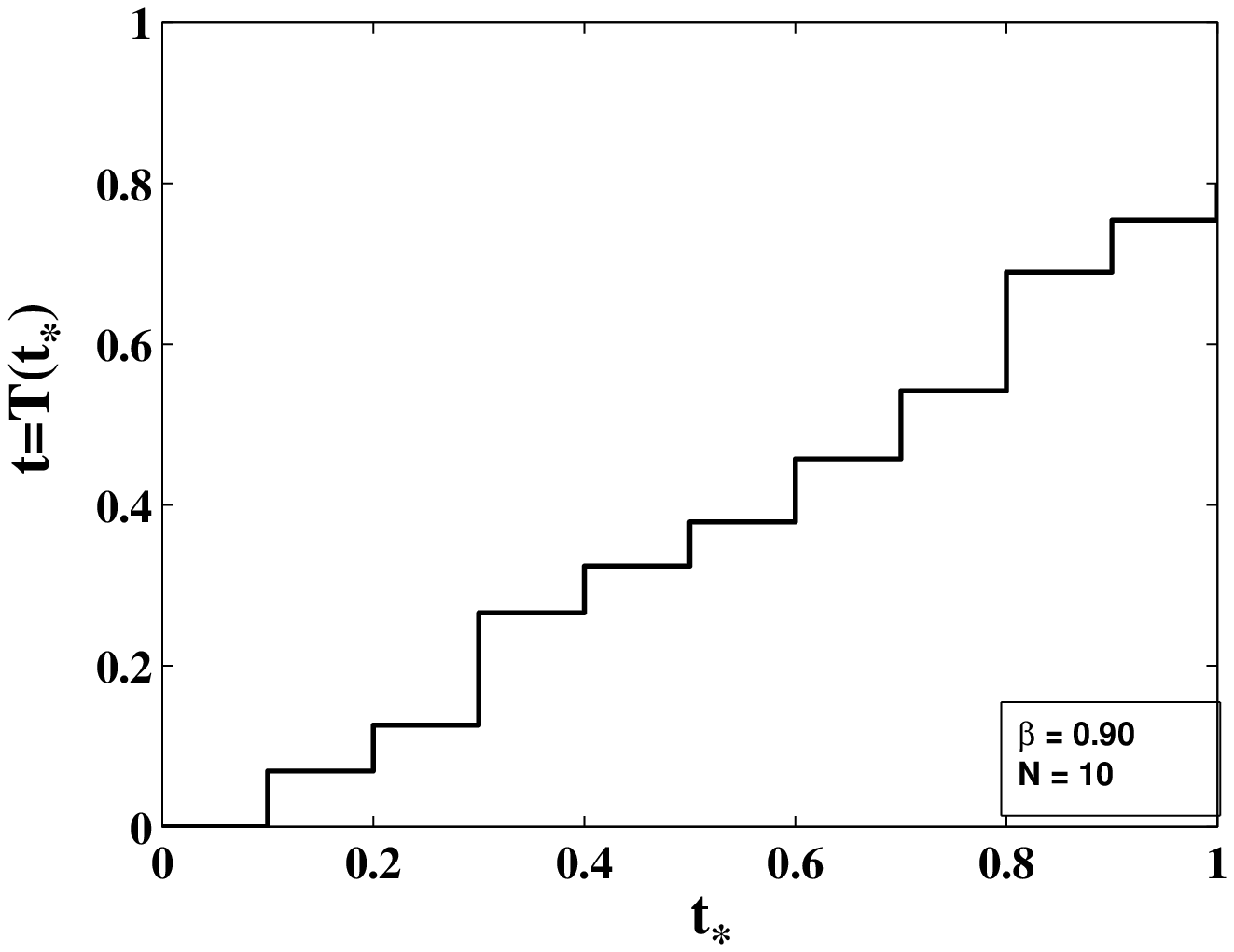}
\includegraphics[width=.52\textwidth]{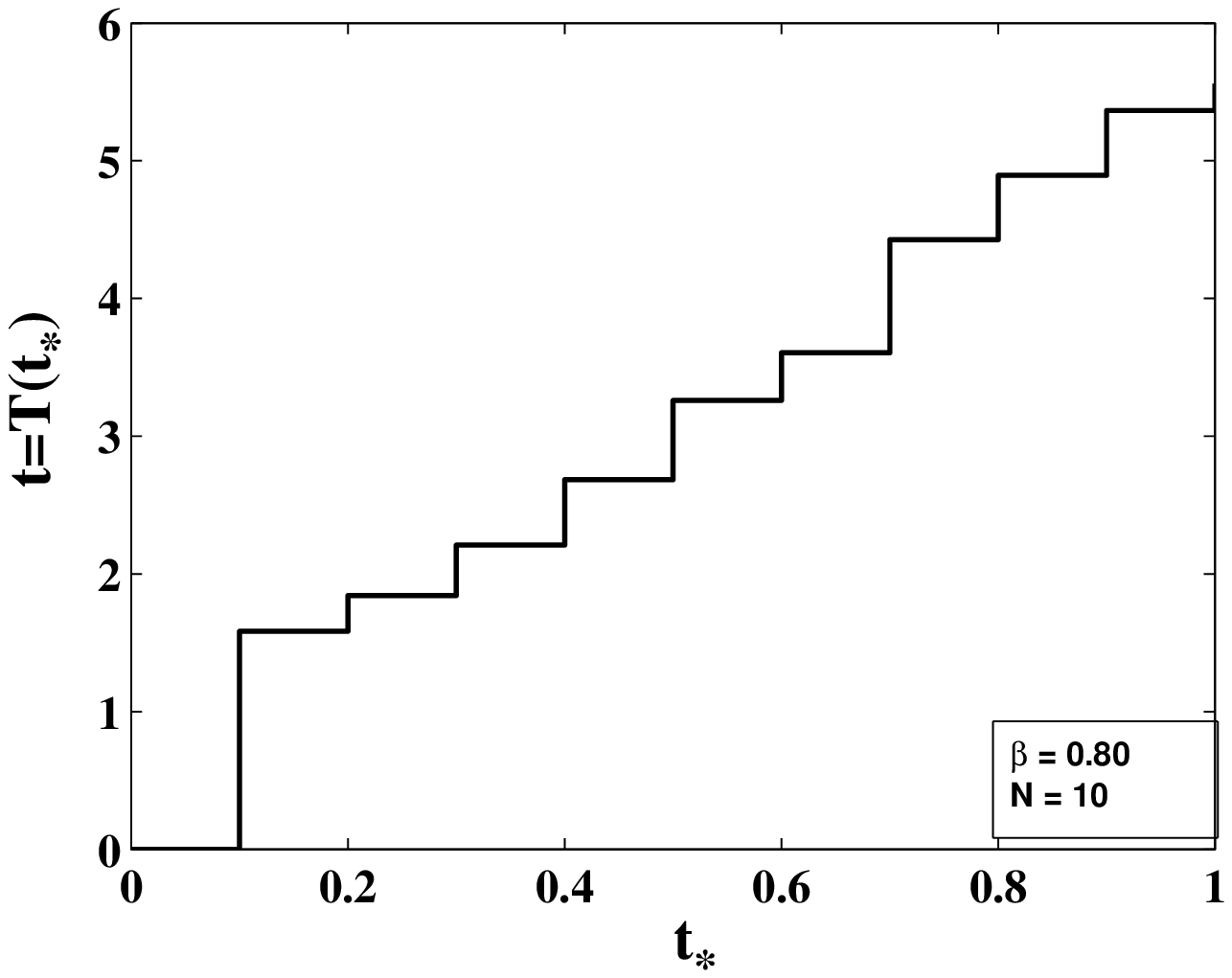}
 \caption{A sample path for the leading process $t=T(t_*)$.}
 \centerline{LEFT: $\{\beta  =0.9,\; N = 10^1 \}$,
      RIGHT: $\{\beta  =0.8,\; N= 10^1 \}$.}

\vskip 0.30truecm
 \includegraphics[width=.52\textwidth]{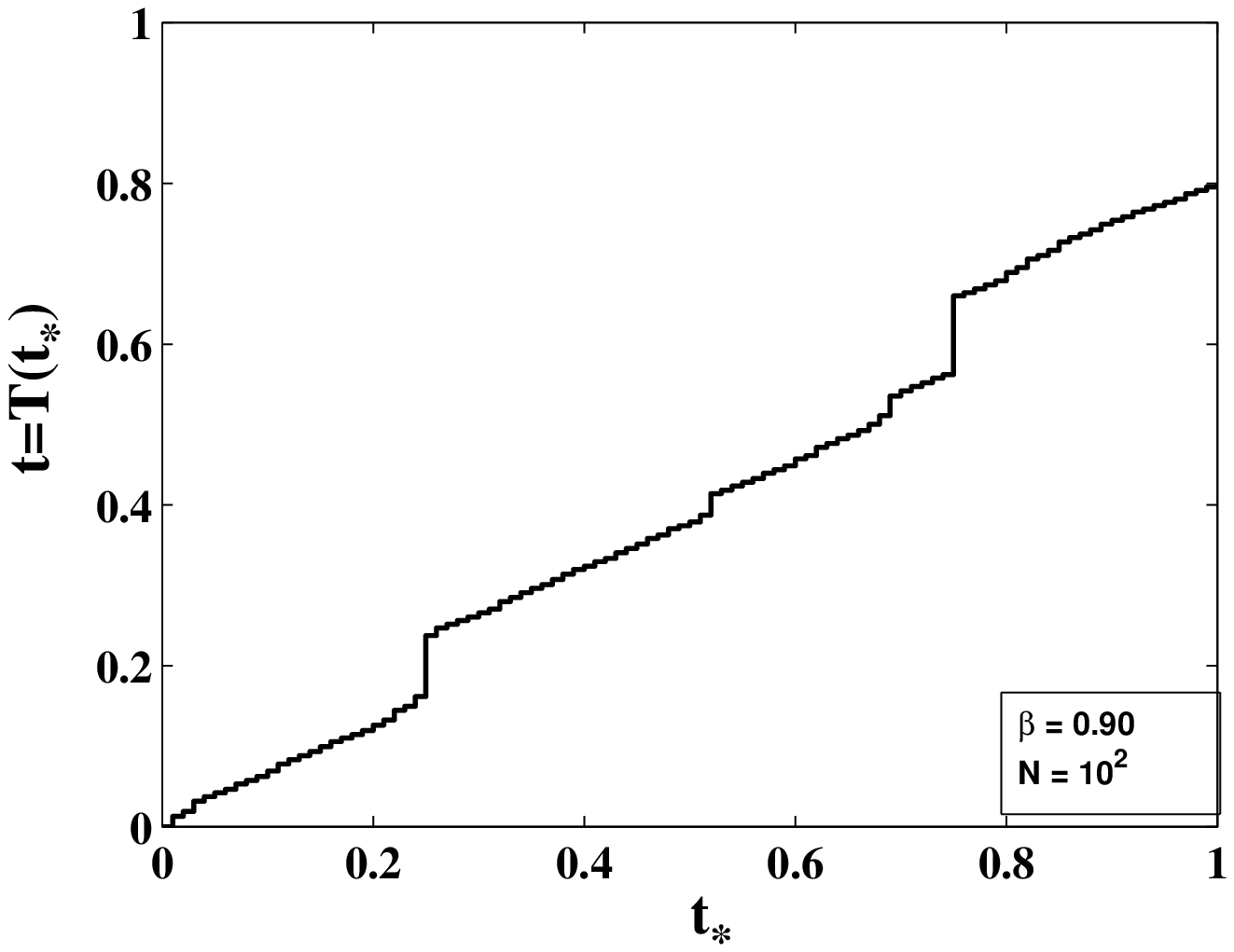}
\includegraphics[width=.52\textwidth]{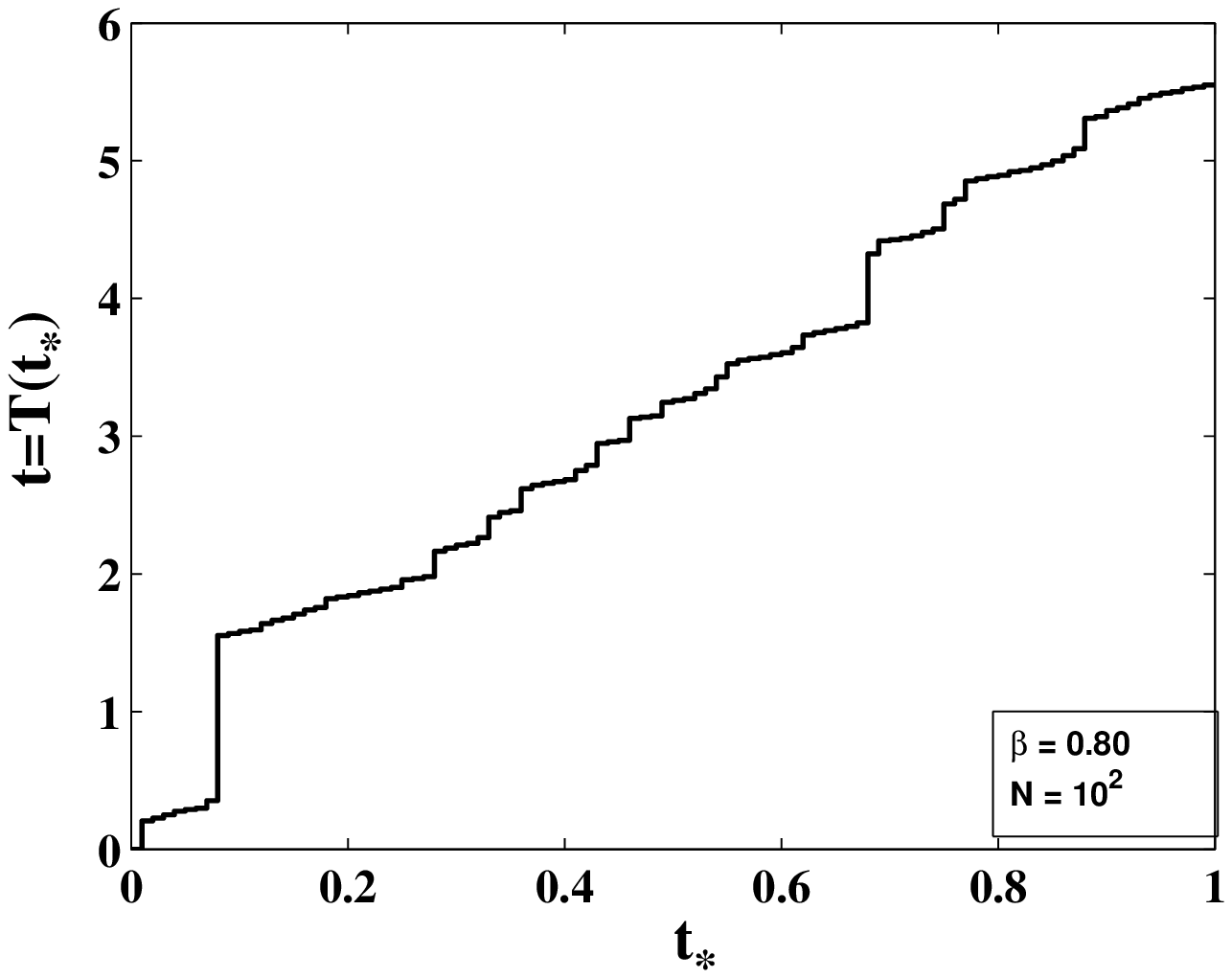}
 \caption{A sample path for the leading process $t=T(t_*)$.}
 \centerline{LEFT: $\{\beta  =0.9,\; N = 10^2 \}$,
      RIGHT: $\{\beta  =0.8,\; N= 10^2 \}$.}

\vskip 0.30truecm
 \includegraphics[width=.52\textwidth]{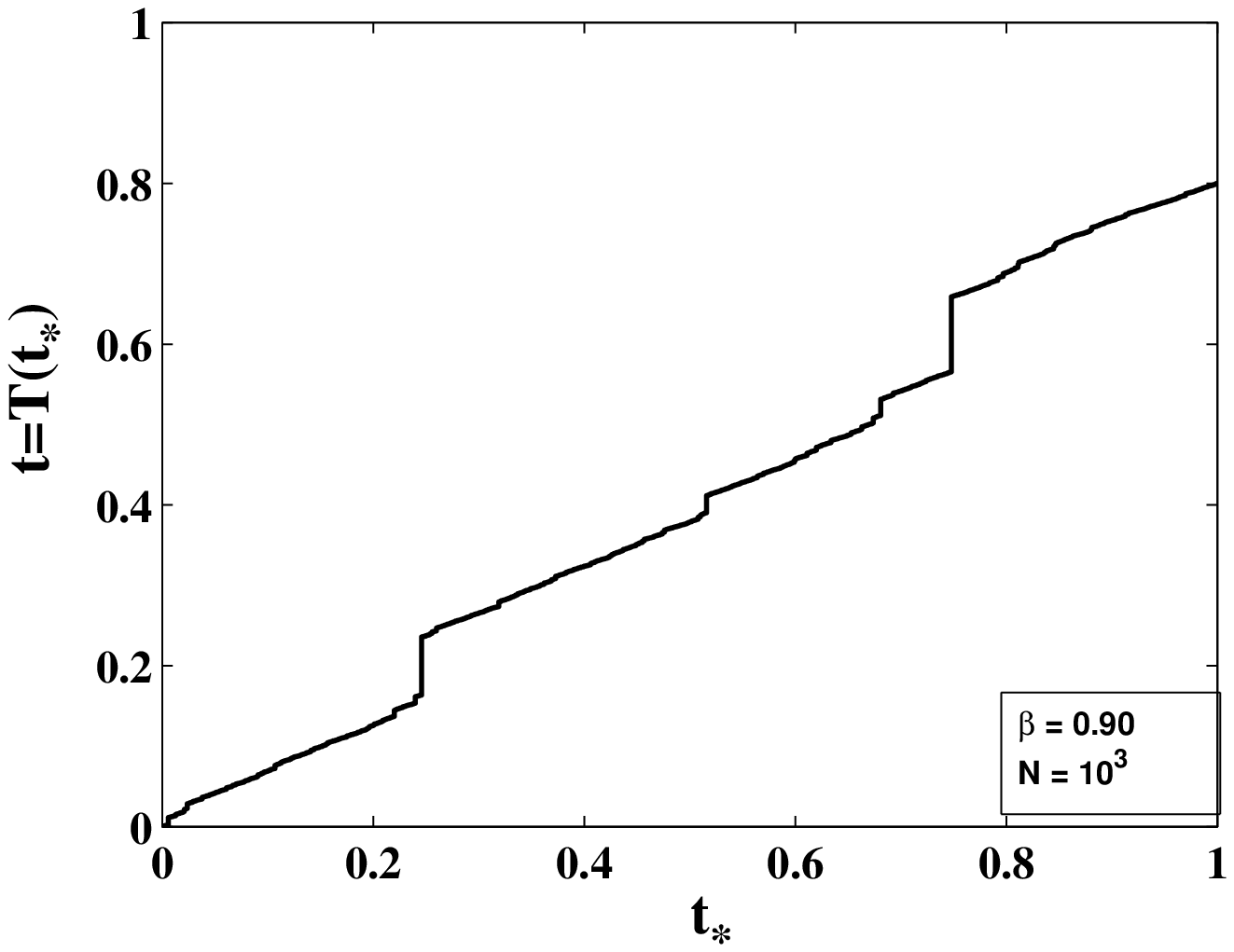}
\includegraphics[width=.52\textwidth]{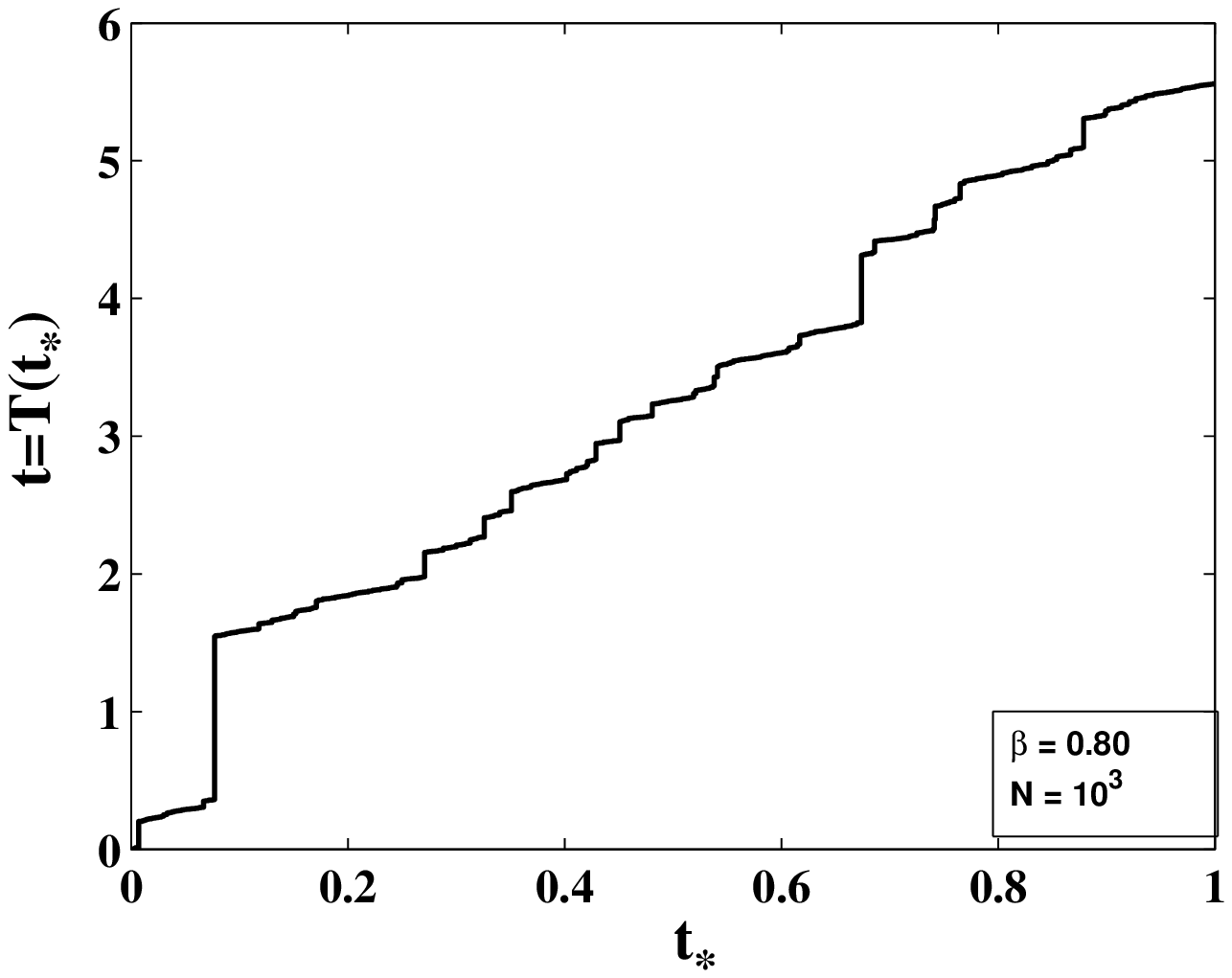}
 \caption{A sample path for the leading process $t=T(t_*)$.}
 \centerline{LEFT: $\{\beta  =0.9,\; N = 10^3 \}$,
      RIGHT: $\{\beta  =0.8,\; N= 10^3 \}$.}
\end{figure}


\begin{figure}
 \includegraphics[width=.52\textwidth]{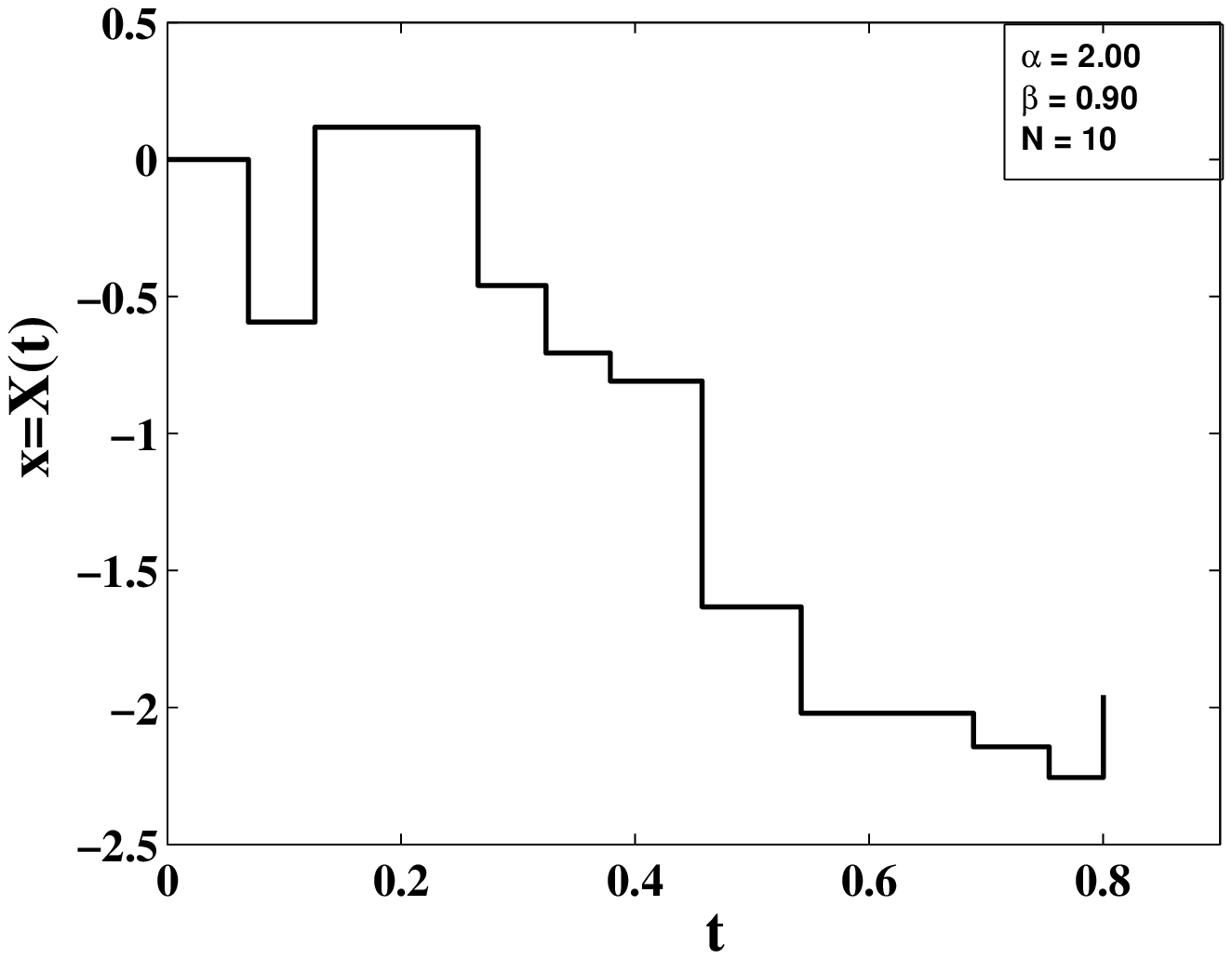}
\includegraphics[width=.52\textwidth]{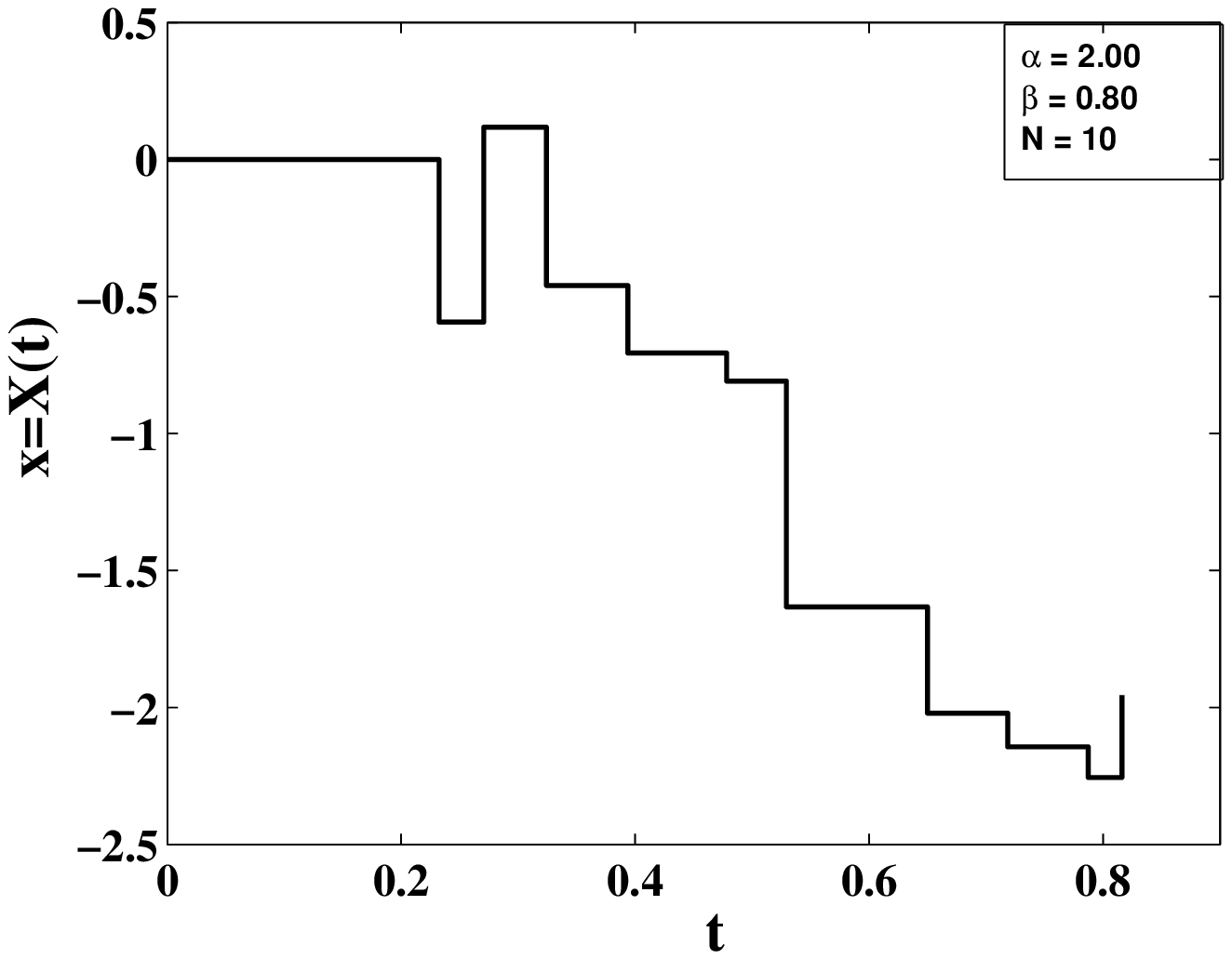}
 \caption{A sample path for the subordinated process $x=X(t)$.}
 \centerline{LEFT: $\{\alpha =2,\; \beta =0.90,\; N=10^1 \}$,
      RIGHT: $\{\alpha =2,\; \beta =0.80,\; N=10^1 \}$.}

\vskip 0.30truecm
 \includegraphics[width=.52\textwidth]{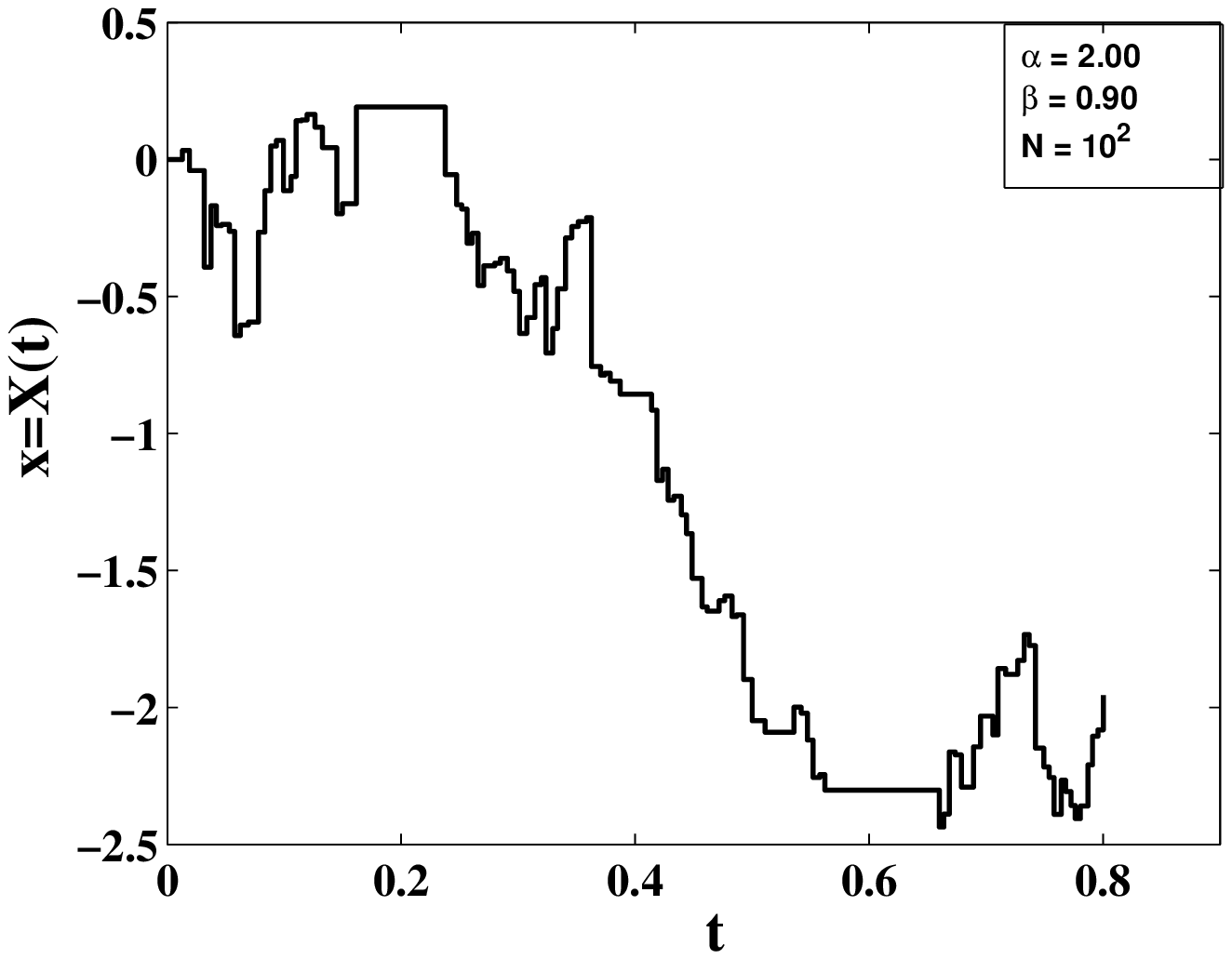}
\includegraphics[width=.52\textwidth]{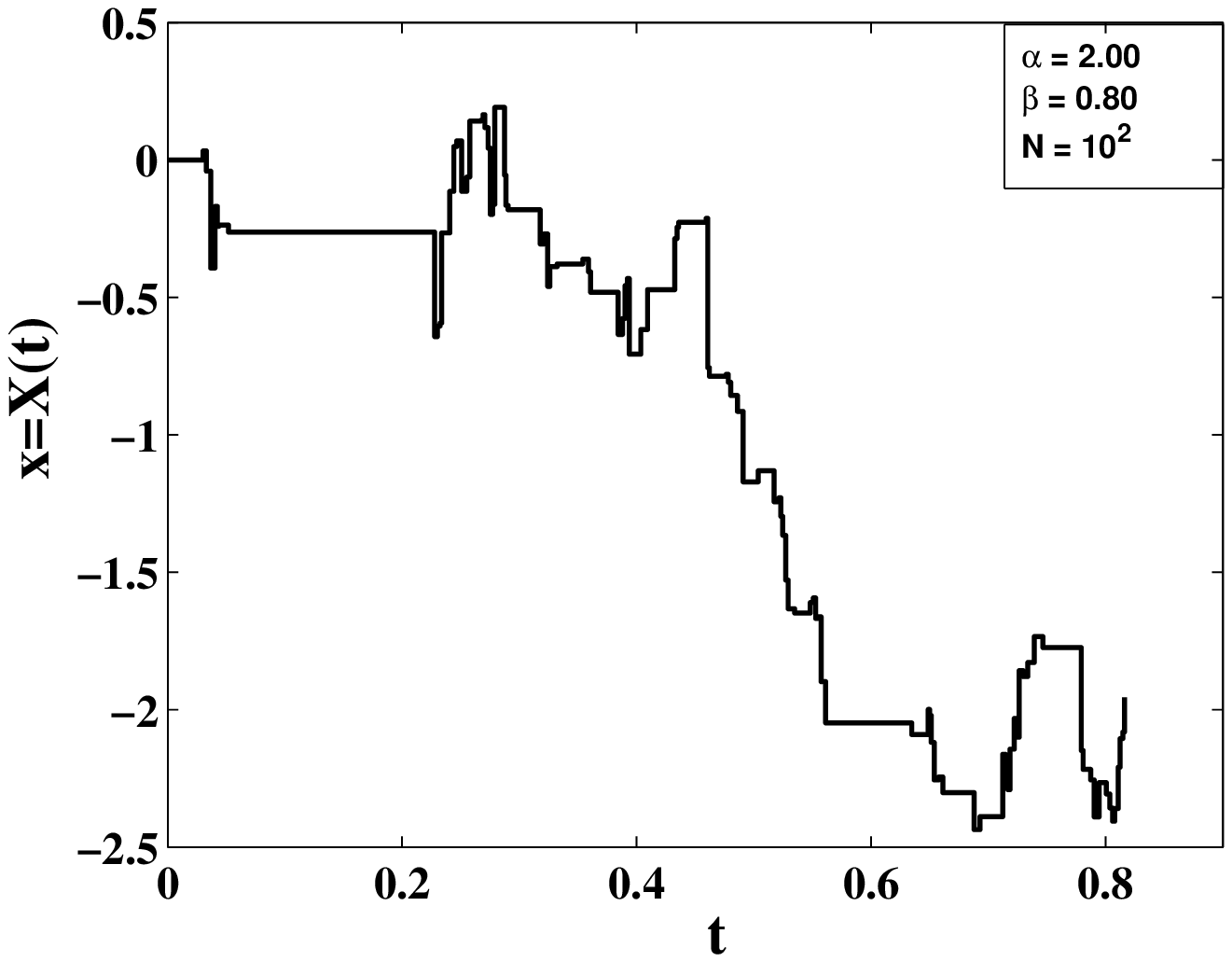}
 \caption{A sample path for the subordinated process $x=X(t)$.}
 \centerline{LEFT: $\{\alpha =2,\; \beta =0.90,\; N=10^2 \}$,
      RIGHT: $\{\alpha =2,\; \beta =0.80,\; N=10^2 \}$.}

\vskip 0.30truecm
 \includegraphics[width=.52\textwidth]{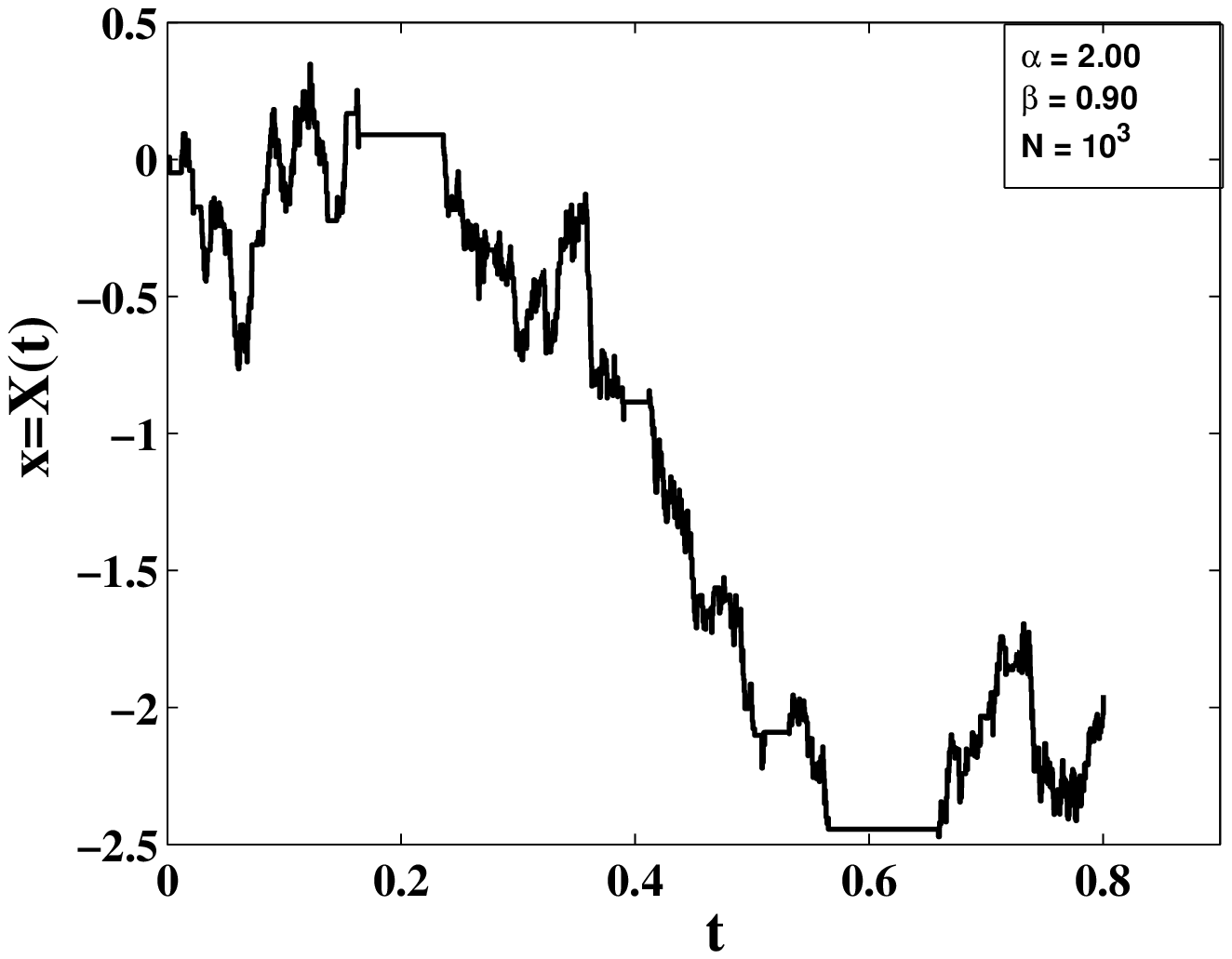}
\includegraphics[width=.52\textwidth]{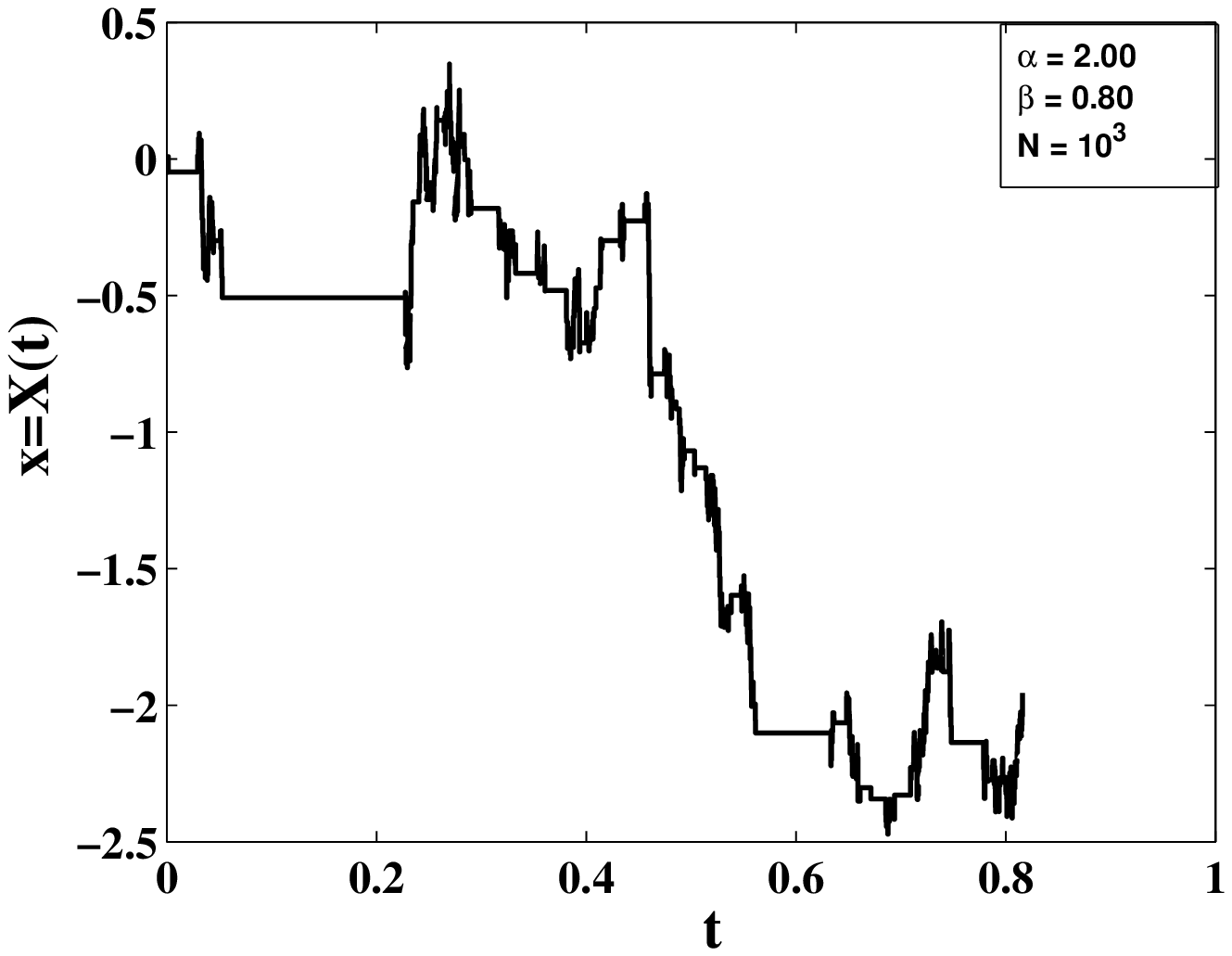}
 \caption{A sample path for the subordinated process $x=X(t)$.}
 \centerline{LEFT: $\{\alpha =2,\; \beta =0.90,\; N=10^3 \}$,
      RIGHT: $\{\alpha =2,\; \beta =0.80,\; N=10^3 \}$.}

\end{figure}

\begin{figure}
 \includegraphics[width=.52\textwidth]{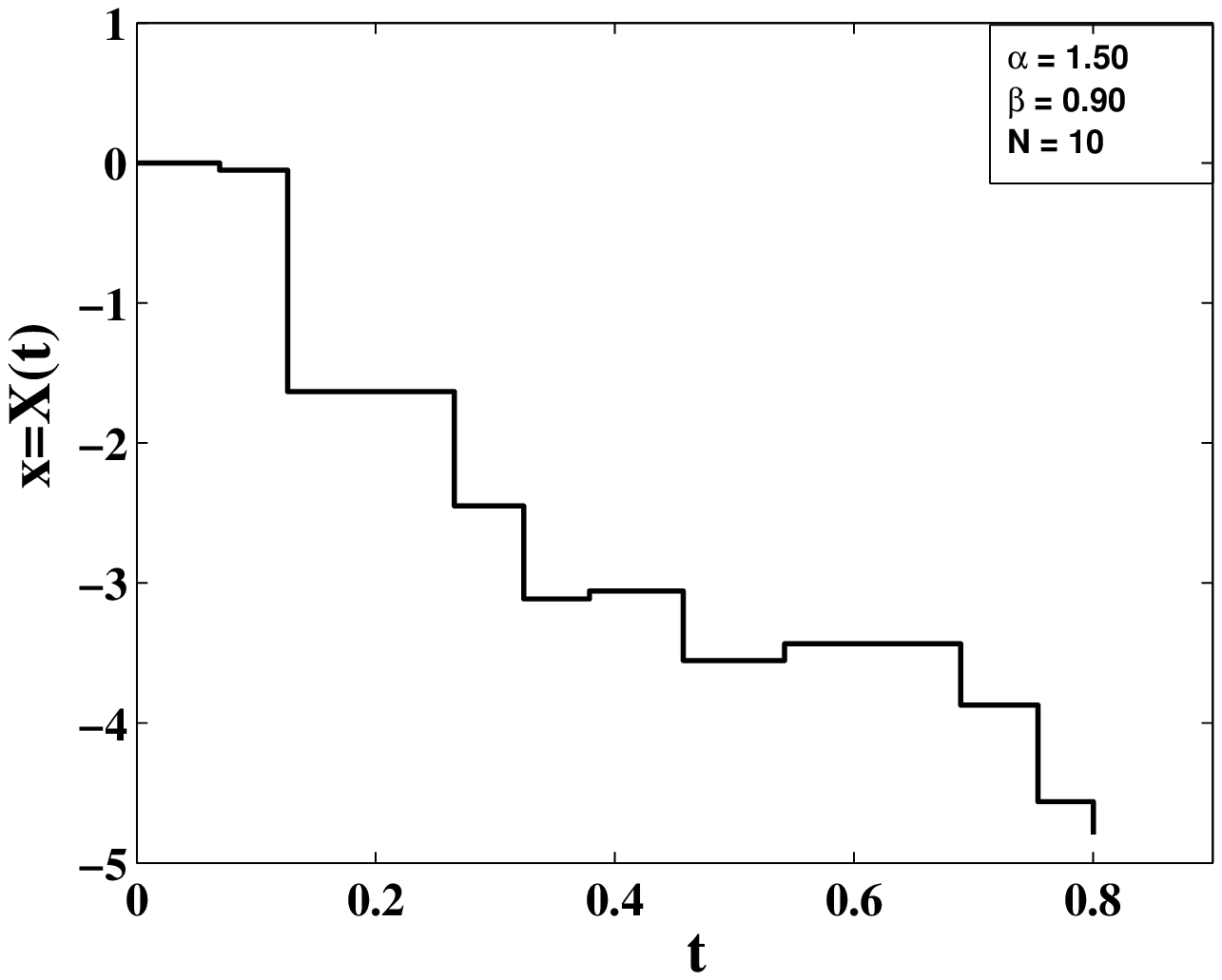}
\includegraphics[width=.52\textwidth]{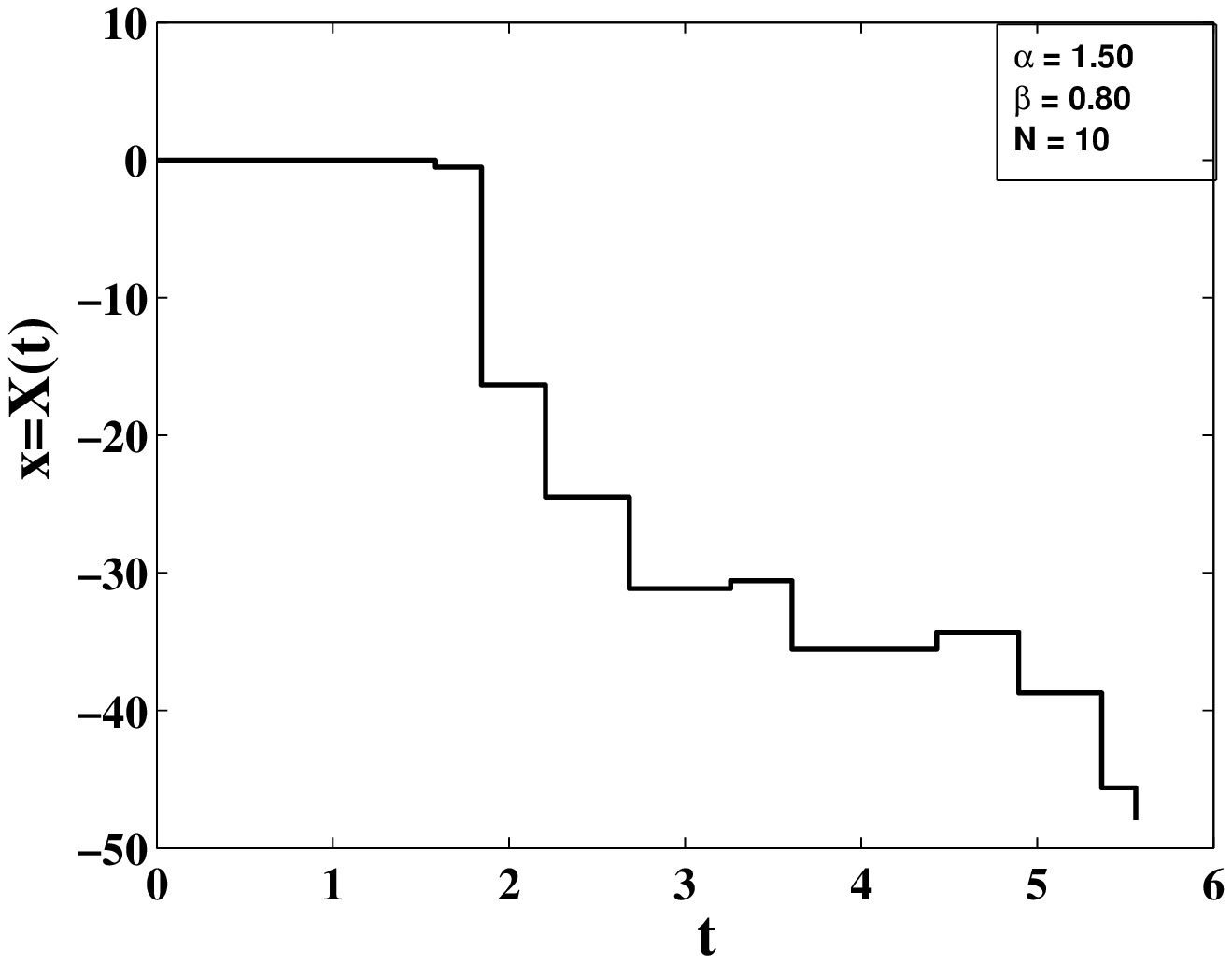}
 \caption{A sample path for the subordinated process $x=X(t)$.}
 \centerline{LEFT: $\{\alpha =1.5,\; \beta =0.90,\; N=10^1 \}$,
      RIGHT: $\{\alpha = 1.5,\; \beta =0.80,\; N=10^1 \}$.}
\vskip 0.30truecm
 \includegraphics[width=.52\textwidth]{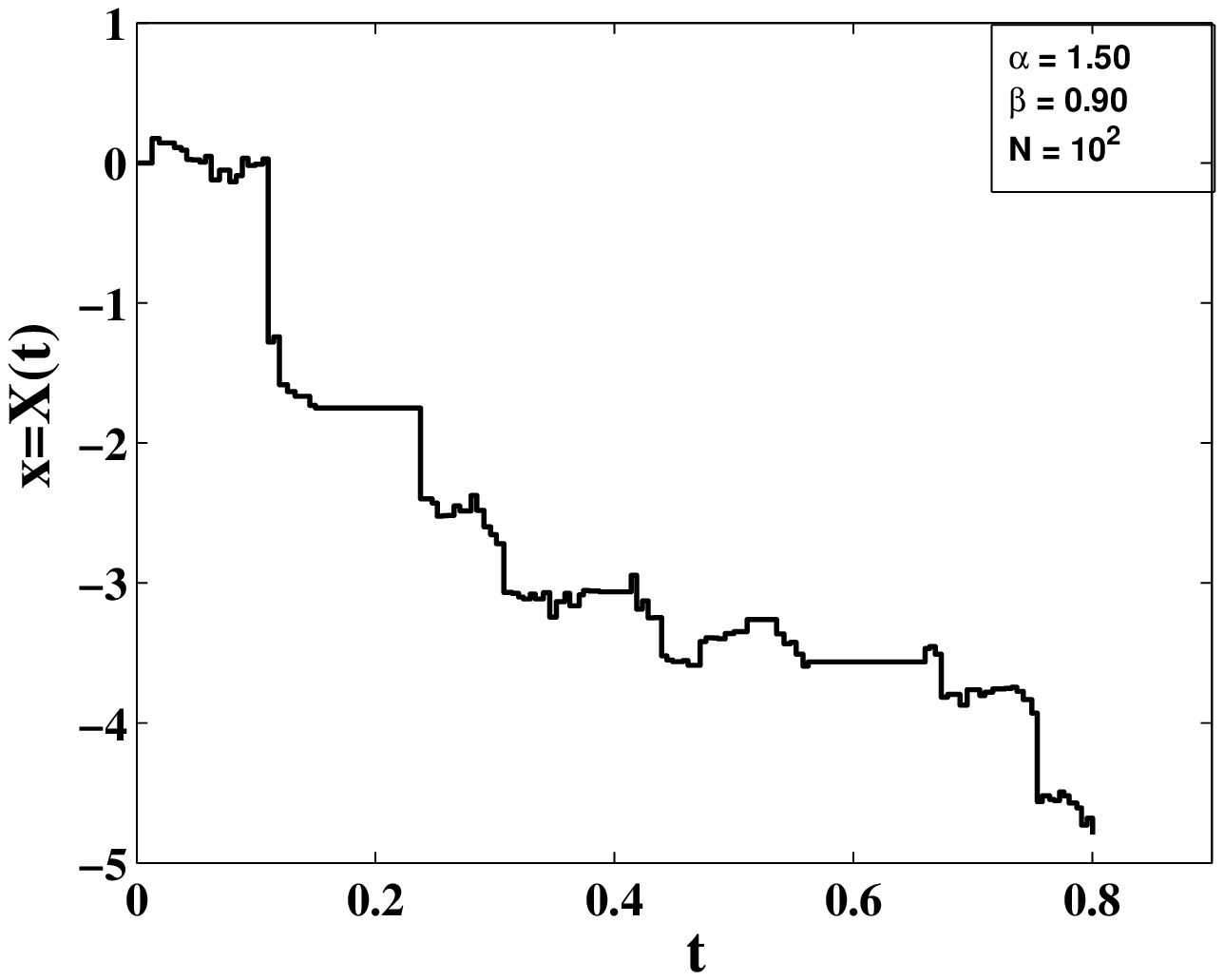}
\includegraphics[width=.52\textwidth]{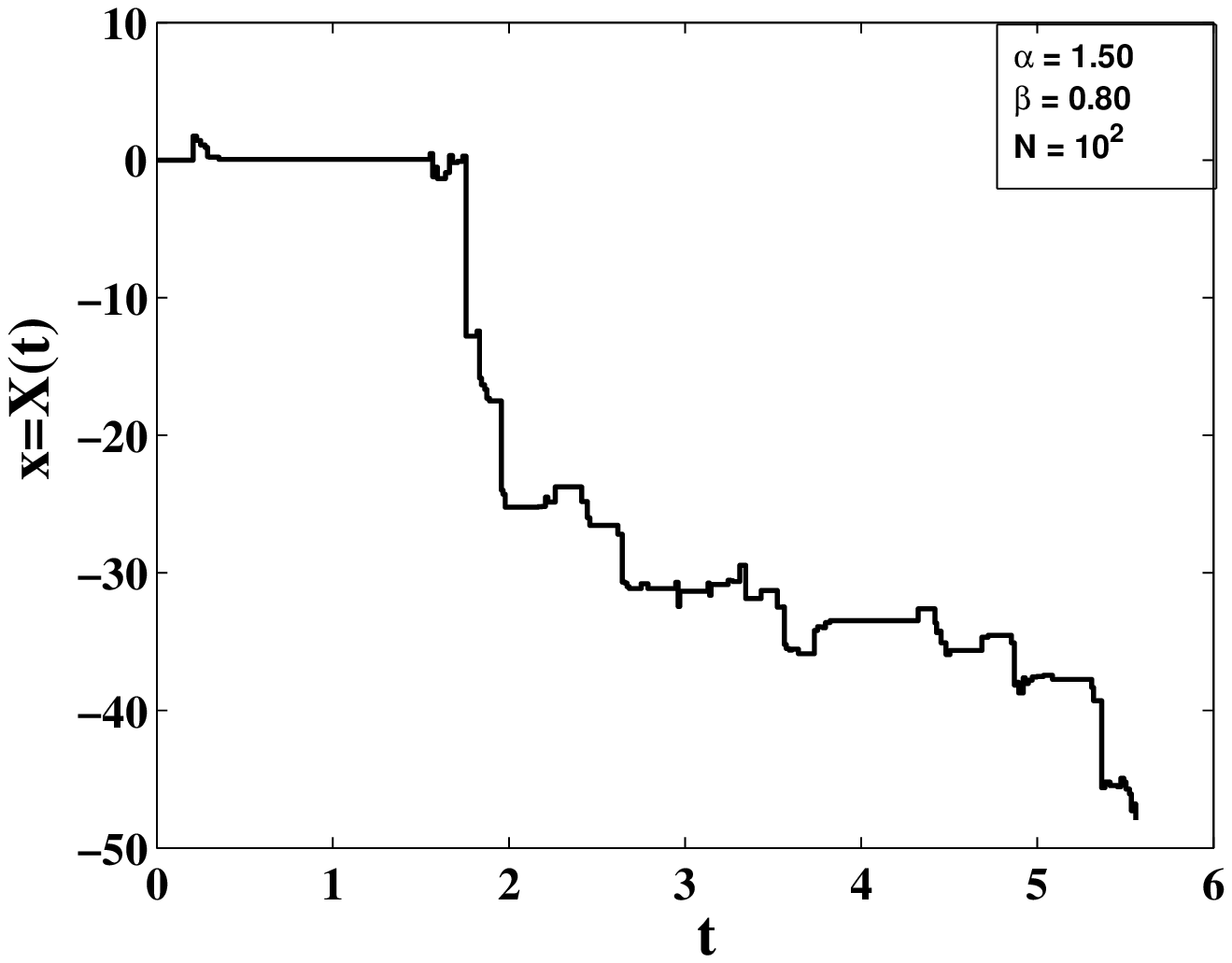}
 \caption{A sample path for the subordinated process $x=X(t)$.}
 \centerline{LEFT: $\{\alpha =1.5,\; \beta =0.90,\; N=10^2 \}$,
      RIGHT: $\{\alpha =1.5,\; \beta =0.80,\; N=10^2 \}$.}
\vskip 0.30truecm
 \includegraphics[width=.52\textwidth]{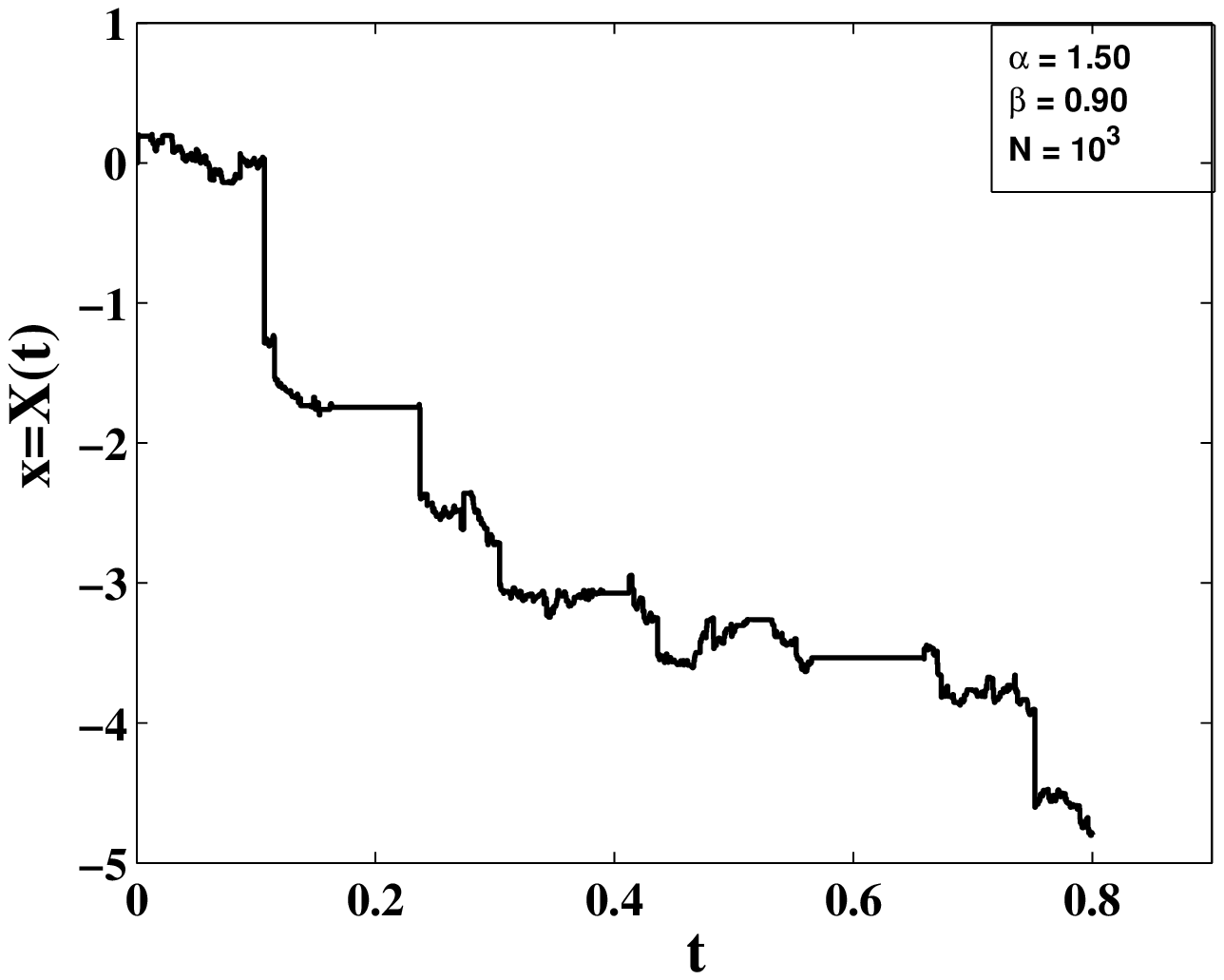}
\includegraphics[width=.52\textwidth]{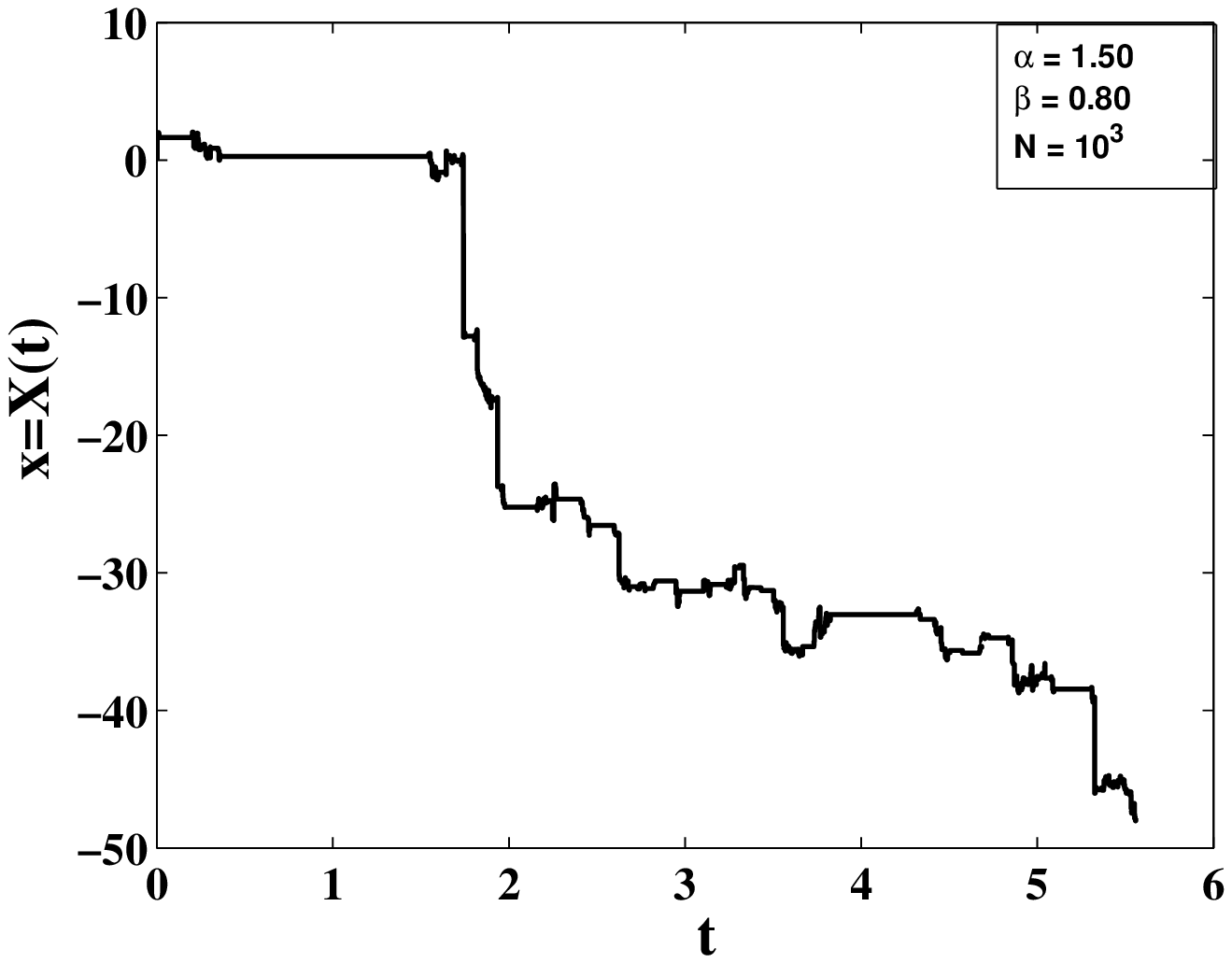}
 \caption{A sample path for the subordinated process $x=X(t)$.}
 \centerline{LEFT: $\{\alpha =1.5,\; \beta =0.90,\; N=10^3 \}$,
      RIGHT: $\{\alpha =1.5,\; \beta =0.80,\; N=10^3 \}$.}
\end{figure}
\section{Conclusions}

Starting from the series
representation (3.5) of the CTRW, 
by considering there the running index of summation as discrete
        operational time and passing to the diffusion limit in a
        well-scaled way, we have shown how to arrive at the integral
        formula of subordination in fractional diffusion.
        Furthermore, we have explained how, in analogy to the
        construction of particle paths in CTRW, particle paths in
        space-time fractional diffusion can be obtained by composition
        of two stable (hence Markovian) processes (one for the
        physical time, the other for the position in space, both
        processes running in operational time). By this composition
        we get in physical space-time the particle path, parametrized
        by the operational time. 
		For this construction of a   particle path we suggest the name 
{\it parametric subordination}.
The essential games are played in
        operational time, for construction of a particle path we avoid
        to   explicitly run the 
        hitting time process (see \cite{M3 PRE02sub})
        generating from physical time the operational time. 
\section*{Acknowledgements}

This work has been  carried out in the framework of a
joint research project for {\it Fractional Calculus
Modelling} ({\tt www.fracalmo.org}).
R.G. and F.M. are grateful to  E. Barkai and M.M. Meerschaert for
inspiring e-mail  exchange of opinions, and
to A.V. Chechkin and I.M. Sokolov for illuminating discussions.





\begin{thebibliography}{99}




\bibitem{BaeumerMeerschaert 01}
 B. Baeumer, M.M. Meerschaert:
 Stochastic solutions for fractional Cauchy problems,
 {\it Fract. Calcul.  Appl. Anal.}
 {\bf 4},  481-500 (2001).


 \bibitem{Balescu 97}
 R. Balescu,
 {\it Statistical Dynamics: Matter out of Equilibrium},
 Imperial College Press - World Scientific, London, 1994,
 Ch. 12, pp. 199-229.
 \bibitem{Barkai PRE01}
 E. Barkai,
 {Fractional Fokker-Planck equation, solution, and application},
 {\it Phys. Rev. E} {\bf 63}, 046118-1/18 (2001).
\bibitem{Barkai ChemPhys02}
E. Barkai,
CTRW pathways to the fractional diffusion equation,
{\it Chem. Phys.} {\bf 284}, 13--27 (2002).
\bibitem{Barkai PRE00}
E. Barkai,  R. Metzler,  J. Klafter,
 From continuous time random walk to fractional
Fokker-Planck equation,
{\em Phys. Rev. E} {\bf 61}, 132-138 (2000).
\bibitem{OEBN EDITOR01}
O.E. Barndorff-Nielsen, T. Mikosch, S.I.  Resnick (Editors),
{\it L\'evy Processes: Theory and Applications},
Birkh\"auser, Boston (2001).
\bibitem{Bochner 55}
S. Bochner,
{\it Harmonic Analysis and the Theory of Probability},
University of California Press, Berkeley (1955).
\bibitem{Bochner 62}
S. Bochner,
 Subordination of non-Gaussian  stochastic processes,
{\em Proc. Nat. Acad. Sciences, USA} {\bf 48}, 19--22 (1962).






\bibitem{Cox RENEWAL67}
D.R. Cox,
{\it Renewal Theory}, Methuen, London (1967). 





\bibitem{Feller 71}
  W. Feller,
 {\em An Introduction to Probability Theory and its Applications\/},
  Vol. 2, Wiley, New York (1971).
\bibitem{GelfandShilov 64}
 I. M. Gel\`{}fand and G. E. Shilov,
 {\it Generalized Functions}, Volume I.
 Academic Press, New York and London (1964).



\bibitem{GAR Vietnam04}
R. Gorenflo and E. Abdel-Rehim,
From power laws to fractional diffusion: the direct way,
{\it Vietnam Journal of Mathematics} {\bf 32} SI,  65-75 (2004).



 \bibitem{GorMai CISM97}
  R. Gorenflo and F. Mainardi,  Fractional calculus:
  integral and differential equations of fractional order,
  in: A. Carpinteri and F. Mai\-nardi (Editors),
  {\em Fractals and Fractional Calculus in Continuum Mechanics\/},
  Springer Verlag, Wien  (1997),  pp. 223--276.
  [Reprinted in {\tt http://www.fracalmo.org}]








\bibitem{GorMai INDIA03}
 R. Gorenflo and F. Mainardi,
 Fractional diffusion processes: probability distributions and
    continuous time random walk, in:
    G. Rangarajan and M. Ding (Editors),
    {\it Processes with Long Range Correlations},
    Springer-Verlag, Berlin (2003), pp. 148-166.
    [Lecture Notes in Physics, No. 621]
\bibitem{GorMai CARRY04}
 R  Gorenflo  and  F. Mainardi,
     Simply and multiply scaled diffusion limits for continuous time
     random walks,
    in: S. Benkadda, X. Leoncini and  G. Zaslavsky (Editors),
   {Proceedings of the  International Workshop on
  Chaotic Transport and Complexity  in Fluids and Plasmas}
     Carry Le Rouet (France) 20-25 June 2004,
     {\it IOP (Institute of Physics) Journal of Physics: Conference
     Series} {\bf 7},  1-16 (2005).




\bibitem{Gorenflo  KONSTANZ01}
 R. Gorenflo, F. Mainardi, E. Scalas and M. Raberto
Fractional calculus and continuous-time finance III:
the diffusion limit,
in:  M. Kohlmann and S. Tang (Editors),
     {\it Mathematical Finance},
    Birkh\"auser Verlag, Basel (2001), pp. 171-180.
\bibitem{GrigoliniRoccoWest 99}
P. Grigolini, A. Rocco and B.J. West,
Fractional calculus as a macroscopic manifestation of randomness,
{\it Phys. Rev. E} {\bf 59}, 2603--2613 (1999).
\bibitem{Hilfer 95a}
   R. Hilfer,                  
  Exact solutions for a class of fractal time random walks,
 {\it  Fractals} {\bf 3}, 211-216 (1995).
\bibitem{Hilfer BOOK}
  R. Hilfer (Editor),
  {\it Applications of Fractional Calculus in Physics},
  World Scientific, Singapore (2000).
 \bibitem{Hilfer PhysA03}
 R. Hilfer,
 On fractional diffusion and continuous time  random walks,
 {\it Physica A}  {\bf 329},  35-39 (2003).
\bibitem{HilferAnton  95}
  R. Hilfer and L. Anton,       
  Fractional master equations and fractal time random walks,
  {\it Phys. Rev. E} {\bf 51}, R848-R851 (1995).
\bibitem{Jacob BOOKS}
N. Jacob,
 {\it Pseudodifferential Operators - Markov Processes},
 {Vol I: Fourier Analysis and Semigroups}
 {Vol II: Generators  and Their Potential Theory},
 {Vol. III: Markov Processes and Applications},
 Imperial College Press,  London  (2001),  (2002), (2005). 
\bibitem{Janicki LN96}
A. Janicki,
{\it Numerical and Statistical Approximation of
Stochastic Differential Equations with Non-Gaussian Measures},
Monograph No 1, H.  Steinhaus Center for Stochastic Methods
in Science and Technology, Technical University,
Wroclaw,  Poland  (1996). 
\bibitem{Janicki-Weron 94}
 A. Janicki and A. Weron,
 {\it Simulation and Chaotic Behavior of
 $\alpha$--Stable Stochastic Processes},
 Marcel Dekker, New York (1994).
\bibitem{Kilbas-et-al BOOK06}          
A.A. Kilbas, H.M. Srivastava  and J.J. Trujillo,
{\it Theory and Applications of Fractional Differential Equations},
Elsevier, Amsterdam (2006).
\bibitem{Kotulski 95a}
M. Kotulski, Asymptotic distributions of continuous-time random walks:
a probabilistic approach,
{\it J. Stat. Phys.} {\bf 81},  777--792 (1995).


\bibitem{Mainardi FCAA01}
 F. Mainardi, Yu. Luchko and G. Pagnini,
     The fundamental solution of the space-time fractional diffusion
     equation,
   {\it Fract. Calcul.  Appl. Anal.}
 {\bf 4}, 153-192 (2001).
 [Reprinted in  {\tt http://www.fracalmo.org}]



\bibitem{Mainardi FCAA03}
 F. Mainardi, G. Pagnini and R. Gorenflo,
 Mellin transform and subordination laws in in fractional diffusion processes,
 {\it   Fract. Calcul.  Appl. Anal.} {\bf 6},  441-459 (2003).

\bibitem{Mainardi JCAM05}
 F. Mainardi, G. Pagnini and R.K. Saxena,
 Fox $H$ functions in fractional diffusion,
 {\it   J. Computat. Appl. Math.} {\bf 178},  321-331 (2005).
\bibitem{Mainardi Bonn00}
   F. Mainardi, M. Raberto, R. Gorenflo and E. Scalas,
 Fractional calculus and continuous-time finance II: the waiting-time
     distribution,
    {\it Physica A} {\bf 287}, 468--481 (2000).
\bibitem{MannellaGrigoliniWest 94}
R. Mannella, P. Grigolini and B.J. West,
A dynamical approach to fractional Brownian motion,
{\it Fractals} {\bf 2}, 81-94 (1994).
\bibitem{M3 PRE02sub}
M.M. Meerschaert, D.A. Benson, H.P  Scheffler and B. Baeumer,
Stochastic solutions of space-fractional diffusion equation,
{\it Phys. Rev. E} {\bf  65},  041103-1/4 (2002).
 \bibitem{M3 PRE02sol}
 M.M. Meerschaert, D.A. Benson, H.P  Scheffler and P. Becker-Kern,
 Governing equations and solutions of anomalous random walk limits,
 {\it Phys. Rev. E} {\bf  66},  060102-1/4 (2002).
\bibitem{Metzler PRE98}
 R. Metzler, J. Klafter, I.M. Sokolov,
 Anomalous transport in external fields:
Continuous time random walks and fractional
diffusion equations extended,
{\em Phys. Rev, E} {\bf 58},  1621-1633 (1998).
\bibitem{Metzler-Klafter PhysRep00}
  R. Metzler and J. Klafter,
 The random walk's guide to anomalous diffusion: a fractional dynamics
 approach, {\it Phys. Reports}  {\bf 339},  1-77 (2000).
 \bibitem{Metzler-Klafter JPhysics04}
 R. Metzler and J. Klafter,
The restaurant at the end of the random walk: Recent developments
 in the description of anomalous transport by fractional dynamics,
 {\it J. Phys. A. Math. Gen.}  {\bf 37},  R161-R208 (2004).
\bibitem{MontrollScher 73}
  E.W. Montroll and H. Scher,
 Random walks on lattices, IV:
 Continuous-time walks and influence of absorbing boundaries,
  {\it J. Stat. Phys.} {\bf 9},  101-135 (1973).
\bibitem{MontrollShles 84}
 E.W. Montroll and M.F. Shlesinger,
  On the wonderful world of random walks, in
  J. Leibowitz and E.W. Montroll (Editors),
  {\it Nonequilibrium Phenomena II: from Stochastics to Hydrodynamics},
 North-Holland, Amsterdam (1984), pp. 1-121.
\bibitem{MontrollWeiss 65}
 E.W. Montroll and G.H. Weiss,
 Random walks on lattices, II,
 {\it J. Math. Phys.}  {\bf 6},  167--181 (1965).
\bibitem{MontrollWest 79}
  E.W. Montroll and  B.J. West,
  On an enriched collection of stochastic processes, in
  E.W. Montroll and J. Leibowitz (Editors),
  {\it Fluctuation Phenomena},
 North-Holland, Amsterdam (1979),  pp. 61-175.
\bibitem{Podlubny 99}
  I. Podlubny,
  {\it Fractional Differential Equations},
  Academic Press, San Diego (1999).
\bibitem{Saichev PhysA05}
A. Piryatinska, A.I. Saichev and W.A. Woyczynski,
 Models of anomalous diffusion: the subdiffusive case,
 {\it Physica A}  {\bf 349},  375-420 (2005).
\bibitem{SaichevZaslavsky 97}
{A. Saichev and G. Zaslavsky},
Fractional kinetic equations: solutions and applications,
{\em Chaos\/} {\bf 7},  753-764 (1997).
\bibitem{SKM 93}
S.G.  Samko, A.A. Kilbas and O.I. Marichev,
{\it Fractional Integrals and Derivatives: Theory  and  Applications},
Gordon and Breach, New York (1993).
\bibitem{Sato 99}
  K-I. Sato,
  {\it L\'evy Processes and Infinitely Divisible Distributions},
  Cambridge University Press, Cambridge (1999).
  \bibitem{Scalas PhysA05}
 E. Scalas,
The application of  continuous-time random walks in finance and economics,
 {\it Physica A} {\bf 362},  225-239 (2006).
\bibitem{SGM 00}
 E. Scalas, R. Gorenflo and F. Mainardi,
 Fractional calculus and continuous-time finance,
 {\it Physica A} {\bf 284},  376-384 (2000).
\bibitem{Scalas PRE04}
 E. Scalas, R. Gorenflo and F. Mainardi,
    Uncoupled continuous-time random walks:
Solution and limiting behavior of the master equation,
{\it Phys. Rev. E} {\bf 69},  011107-1/8 (2004).
\bibitem{ShlesZaslKlafter Nature93}
 M.F. Shlesinger, G.M. Zaslavsky, and J. Klafter,
Strange kinetics,
 {\it Nature} {\bf 363},  31-37 (1993).
 \bibitem{Sokolov PRE01a}
I.M. Sokolov,
L\'evy flights from a continuous-time process,
 {\it Phys. Rev. E} {\bf 63} 011104-1/10 (2001).
\bibitem{Sokolov PRE01b}
 I.M. Sokolov,
Thermodynamics and  fractional Fokker-Planck equation,
{\it Phys. Rev. E} {\bf 63},  056111-1/8 (2001).
\bibitem{Sokolov PRE02}
I.M. Sokolov,
Solutions of a class of non-Markovian Fokker-Planck equations,
 {\it Phys. Rev. E} {\bf 66} 041101-1/5 (2002).
 \bibitem{Sokolov PhysA01}
 I.M. Sokolov,  J. Klafter,  A. Blumen,
 Linear response in complex systems: CTRW and the fractional
Fokker-Planck equations,
{\em Physica A} {\bf 302}, 268-278 (2001).
\bibitem{SokolovKlafter EINSTEIN05}
 I.M. Sokolov and J. Klafter,
From diffusion to anomalous diffusion: a century after Einstein's
Brownian motion,
{\it Chaos} {\bf 15}, 026103-026109 (2005).
\bibitem{Sokolov-Klafter-Blumen_PhysicsToday02}
I.M. Sokolov,  J. Klafter and A. Blumen,
Fractional kinetics,
{\it Physics Today} {\bf 55},  48-54 (2002).
\bibitem{Stanislavsky PRE00}
A.A. Stanislavski,
Memory effects and  macroscopic manifestation of randomness,
{\it Phys. Rev. E} {\bf 61},  4752--4759 (2000).
\bibitem{Stanislavsky PHYSA03}
A.A. Stanislavsky,
Black-Scholes model under subordination,
\emph{Physica A} {\bf 318}, 469-474 (2003).
\bibitem{UchaikinSaenko 03}
 V.V. Uchaikin and V.V. Saenko,
 Stochastic solution of partial differential equations
 of fractional orders,
 {\it Siberian Journal of Numerical Mathematics} {\bf 6},
 197-203 (2003).
\bibitem{UchaikinZolotarev 99}
{V.V. Uchaikin and V.M. Zolotarev},
 {\it Chance and Stability. Stable Distributions and their Applications},
 VSP, Utrecht (1999).
\bibitem{Weiss BOOK94}
 G.H. Weiss,
 {\it Aspects and Applications of Random Walks},
 North-Holland, Amsterdam (1994).
\bibitem{West BOOK03}
B.J. West, M. Bologna and P. Grigolini.
{\it Physics of Fractal Operators}, Springer Verlag, New York (2003).


 \bibitem{Wyss-Wyss 01}
 {M.M. Wyss  and  W. Wyss},
 Evolution, its fractional extension and generalization,
{\it Fract. Calcul.  Appl. Anal.} {\bf 4}, 273-284 (2001).
 \bibitem{Zaslavsky PhysRep02}
 G.M. Zaslavsky,
 Chaos,  fractional kinetics  and anomalous transport,
{\it Phys. Reports}  {\bf 371},  461-580 (2002).
\bibitem{Zaslavsky BOOK05}
G.M. Zaslavsky.
{\it Hamiltonian Chaos and Fractional Dynamics},
Oxford University Press, Oxford (2005).
\end{thebibliography}
\end{document}